\documentclass[twocolumn]{aastex631}

\newcommand{\noop}[1]{}

\shorttitle{Devouring The Centaurus~A Satellites}
\shortauthors{Weerasooriya et al.}
\graphicspath{{./}{}}
\begin{document}

%\title{Simulating Dandelions of the Milky Way: Using \textsc{Galacticus} to Model Dwarf Satellites of the Milky Way}

\title{Devouring The Centaurus~A Satellites: Modeling Dwarf Galaxies with \textsc{Galacticus}}

\author[0000-0001-9485-6536]{Sachi Weerasooriya}
\affiliation{Department of Physics and Astronomy, Texas Christian University, Fort Worth, TX 76109, USA}
\affiliation{Carnegie Observatories, 813 Santa Barbara Street, Pasadena, California, 91101 USA}

\author[0000-0003-4037-5360]{Mia Sauda Bovill}
\affiliation{Department of Astronomy, University of Maryland, College Park, MD 20742}
\affiliation{Department of Physics and Astronomy, Texas Christian University, Fort Worth, TX 76109, USA}

\author{Matthew A. Taylor}
\affiliation{ University of Calgary,
2500 University Drive NW,
Calgary Alberta T2N 1N4,
CANADA}

 \author[0000-0001-5501-6008]{Andrew J. Benson}
 \affiliation{Carnegie Observatories, 813 Santa Barbara Street, Pasadena, California, 91101 USA}

\author{Cameron Leahy}
\affiliation{ University of Calgary,
2500 University Drive NW,
Calgary Alberta T2N 1N4,
CANADA}

\begin{abstract}
 
For the first time, systematic studies of dwarf galaxies are being conducted throughout the Local Volume, including the dwarf satellites of the nearby giant elliptical galaxy Centaurus~A (NGC 5128). Given Centaurus~A’s mass (roughly ten times larger than that of the Milky Way), AGN activity, and recent major mergers, investigating the dwarf galaxies of Centaurus~A and their star formation physics is imperative.  However, simulating the faintest dwarfs around a galaxy of Centaurus~A’s mass with sufficient resolution in a hydrodynamic simulation is computationally expensive and currently infeasible. In this study, we seek to reproduce the properties of Centaurus~A dwarfs using the semi-analytic model \textsc{Galacticus} to model dwarfs within a 700 kpc region around Centaurus A, corresponding approximately to its splashback radius. We investigate the effects of host halo mass and environment and predict observable properties of Centaurus~A dwarfs using astrophysical prescriptions and parameters previously tuned to match properties of the Milky Way's satellite galaxies. This approach allows us to approximately replicate cumulative luminosity functions, and luminosity--metallicity and luminosity--half-light-radii relations observed in the Centaurus~A satellites. We provide predictions for the velocity dispersions, and star formation histories of Centaurus~A dwarfs. The agreement between our predicted star formation histories for Centaurus~A dwarfs and those of the Milky Way dwarfs implies the presence of universal processes governing star formation in the dwarf galaxies. Overall, our findings shed light on the star formation physics of dwarf galaxies in the Centaurus~A system, revealing insights into their properties and dependence on the host environment.

 % \textbf{} The agreement between our predicted SFHs for Centaurus~A dwarfs and those of the Milky Way implies the presence of universal processes governing star formation in \textbf{the overall sample of galaxies}. Overall, our findings shed light on the star formation physics of dwarf galaxies in the Centaurus~A system, revealing insights into their properties and dependence on the host environment.
\end{abstract}

\section{Introduction}\label{sec:intro}

Dwarf galaxies are the fundamental building blocks of larger structures. They are the most abundant type of galaxies in the universe at all redshifts \citep[{\it e.g.},][]{Binggeli+1988,ferguson1994,Marzke&daCosta1997}, and the dwarf galaxies of the Local Group are well studied both observationally \citep{sdss2014,SDSS,pandas2016,DES,mcconnachie2012,York+2012} and theoretically \citep{FIRE2018,Applebaum2021,sommerville2020,Shipp+2022,Weerasooriya+2022}.

The shallow gravitational potential wells of dwarf galaxies make them extremely sensitive to both internal feedback and external, environmental processes \citep[{\it e.g.},][]{dekel1986,thoul1996,benson2002a,okamoto2010}. This quality makes them excellent objects with which to study the effects of the surrounding environment and its role in dwarf galaxy physics. Previous studies have shown that dwarf galaxies are susceptible to quenching of their star formation through various processes including ram pressure stripping \citep[e.g. process of stripping of gas from an infalling satellite galaxy by the gaseous halo of the host galaxy; ][]{Grebel+2003,Gorkom+2004,Tonnesen&Bryan2009}, stellar feedback, and tidal stripping. Among these quenching mechanisms, ram pressure stripping has been shown to be the primary mechanism of quenching in simulations \citep{Murakami&Babul1999,Mayer+2006,Slater&Bell2014, Kazantzidis+2017, Simpson+2018}, while stellar feedback can increase the efficiency of ram pressure stripping by heating the gas reservoir \citep{Bahe&McCarthy2015,Kazantzidis+2017}.
The ELVES III survey \citep{Greene+2023} has explored  quenching fractions of quiescent and star forming galaxies for 30 Local Volume hosts, finding that quenching fractions in these hosts are similar to that of the Milky Way, and that the quenching time increase with stellar mass. ~\cite{akins2021} found that galaxies with stellar masses $M_*\geq 10^8 \,\mathrm{M}_{\odot}$ show resistance to environmental quenching.\\

To assess the effectiveness of ram pressure stripping, \cite{Taibi+2022} examined metallicity gradients in galaxies as a potential signature of radially-dependent environmental quenching. In particular, the gas content of a galaxy is expected to be directly linked to ram pressure stripping, where galaxies with lower gas content are more likely to have undergone this process. However, \cite{Taibi+2022} found no correlation of metallicity gradient with internal gas content or distance from the host galaxy. \cite{Taibi+2022} categorized their galaxy sample based on HI gas content but did not observe a statistically significant difference between gas-poor and gas-rich systems. These findings differ from the results reported by \cite{2013ApJ...767..131L}. In their study, gas-poor systems exhibited a radially decreasing metallicity profile, while gas-rich systems displayed a flat profile. \cite{Taibi+2022} attribute this contradiction primarily to a difference in sample sizes. Their results suggest a limited role of environmental interactions with satellite galaxies in shaping their metallicity gradients. Although an increase in the scatter of metallicity gradients was observed, its statistical insignificance makes it challenging to confirm whether ram pressure (or tidal) stripping plays a significant role in influencing metallicity gradients. Given the sensitivity of dwarf galaxies to these quenching effects, it is important to understand how they affect dwarf galaxy properties such as their luminosities and their metal content, and how such effects depend on the environment and mass of the host halo.\\

While studies of dwarfs in the Local Group alone cannot give a clear picture of the role environment plays in the physics of their evolution, several studies have now explored Milky Way-like galaxies in the Local Volume \citep{Geha+2017,Mao+2021,Carlsten+2022}. Such observational studies of the Local Volume include the ELVES survey \citep{Greene+2023}, and surveys of cluster scale environments such as Virgo $\sim10^{15}\,\mathrm{M}_{\odot}$ and Fornax $\sim10^{14}\,\mathrm{M}_{\odot}$ \citep{mcconnachie2012,Richardson+2011,Ferrarese+2012,Eigenthaler+2018}.  With the upcoming unprecedented volume of observations through soon-to-be available facilities such as the Nancy Grace Roman telescope and Rubin Observatory, it is essential that we explore dwarfs beyond the Local Group with theoretical models. One method of investigation is through the luminosity-metallicity relation, which is imprinted with the stellar feedback physics of dwarf galaxies. Studies have shown that the mass--metallicity relationships of both the Milky Way and M31 follow similar trends \citep{Kirby+2020}. Thus, it is imperative that we look beyond the Local Group (LG) to understand whether or not properties in dwarf galaxies are independent of the environment, specifically the mass of their host halo \footnote{To clarify terminology, we refer to the, typically low-mass, halo in which a satellite dwarf galaxy formed, and which it resides at the center of, as its ``subhalo''. We refer to the halo of the larger galaxy within which the dwarf satellite orbits as its ``host halo''}. \\

Centaurus~A is the closest, easily observable giant elliptical galaxy located 3.8~Mpc from the Milky Way \citep{Harris+2010}. It provides the most accessible opportunity to study dwarf galaxies within a higher mass host and in an environment which sits between group and cluster scales. The question of how dwarf galaxy physics are affected by different mass hosts needs further exploration in these intermediate mass host environments. In recent years the halo of Centaurus~A has been targeted by several surveys including the Survey of Centaurus~A's Baryonic Structures \citep[SCABS;][]{SCABS,Taylor+2017,Taylor+2018} and the Panoramic Imaging Survey of Centaurus \& Sculptor \citep[PISCeS;][]{Sand+2014,crnojevic2014,crnojevic2016}.  These are the first systematic surveys of the dwarf satellites of Centaurus~A. In addition, studies by \cite{muller2015,muller2017,Muller2019,Muller+2021,Muller+2022} have also covered both M83 and Centaurus~A with DECam. As a result of these surveys the number of known or suspected dwarf galaxies has almost doubled over the past 5--10 years.\\

The mass of Centaurus~A is currently not well constrained. Several studies have attempted to estimate the virial mass of Centaurus~A using a variety of methods. For example, \cite{woodley2007} estimated the mass of Centaurus~A (specifically the pressure supported and rotational supported mass) using the globular cluster population within $50$~kpc (finding a mass of $1.3\times 10^{12}\,\mathrm{M}_{\odot}$). \cite{van2000} calculated the mass of Centaurus~A using the virial theorem ($1.4\times 10^{13}\,\mathrm{M}_{\odot}$) and the projected mass method out to $640$ kpc ($1.8\times 10^{13}\,\mathrm{M}_{\odot}$). \cite{Muller+2022} estimates a dynamical mass of $1.2\times 10^{13}\,\mathrm{M}_{\odot}$ within $800$~kpc. \cite{Pearson+2022} has placed a lower limit ($> 4.7\times10^{12}\,\mathrm{M}_{\odot}$) on the mass of Centaurus~A using stellar stream models. The upper range of virial masses measured for Centaurus~A is $\sim10^{13} \mathrm{M}_{\odot}$ \citep{van2000,peng+2004,woodley2007,lokas,Harris+2015}, which falls between the masses of Milky Way and large clusters such as Virgo and Fornax.\\

To date however, few theoretical studies of the Centaurus~A system have been carried out. \cite{bovill2016} used a high-resolution N-body simulation that did not include baryons to study the Centaurus~A globular clusters, and \cite{Muller2019} compared the luminosity function within 200~kpc to Centaurus~A analogs in the TNG100 simulation \citep{Pillepich+2018}. However, hydrodynamical simulations of massive systems outside our Galactic neighborhood that have been run all the way to $z\,=\,0$ do not meet the resolution required to fully resolve the star formation physics of fainter dwarfs. For example, TNG100 has a resolution of $m_\mathrm{DM}=7.5\times 10^6\,\mathrm{M}_{\odot}$ and $m_{\mathrm{baryon}}=1.4\times 10^6\,\mathrm{M}_{\odot}$ resolving galaxies with $M_*>10^8\,\mathrm{M}_{\odot}$ \citep{Pillepich+2018}. This is barely sufficient to resolve Small Magellanic Cloud analogs and is inadequate to resolve fainter dwarfs. Even the higher resolution TNG50 simulation \citep{Nelson+2019} can only resolve dwarf galaxies down to $M_*=10^8\,\mathrm{M}_{\odot}$. Therefore, we need to explore more computationally efficient techniques, such as semi-analytic models (SAMs) that provide an efficient method to explore the star formation physics of dwarf galaxies.\\
    
Our previous work modeling the satellites of the Milky Way \citep{Weerasooriya+2022} made use of the \textsc{Galacticus} SAM \citep{galacticus}, and demonstrated that it can be constrained to simultaneously and accurately reproduce the luminosity function and luminosity--metallicity relation of Milky Way dwarfs, while also successfully reproducing the velocity dispersions, half-light radii, mass-to-light ratios, and star formation histories of dwarf galaxies down to and including the SDSS ultra faints. The key baryonic physics in this model includes reionization occurring at $z=9$ (and subsequently heating the intergalactic medium, preventing accretion of gas into halos with virial velocities below $25\;\mathrm{km/s}$), and star formation following the model of \cite{Blitz&Rosolowsky2006}. In our SAM, stellar feedback accounted for by assuming that stellar winds and supernovae driven outflows remove gas from the interstellar medium of a galaxy, with this mass loss becoming inefficient in galaxies with circular velocities above approximately $150\,\mathrm{km/s}$, and becoming more efficient in shallower potential wells. Both ram pressure and tidal stripping of baryons is also included. Full details of the model can be found in \cite{Weerasooriya+2022}---we will also review key components of the model relevant to this work in \S\ref{sec:sam}. In this work, we run \textsc{Galacticus} on two different sets of halo merger trees: one generated using extended Press-Schechter (EPS) theory, and a second taken from an $N$-body simulation of a Centaurus~A analog. We use the same astrophysical models (as described in detail in \cite{Weerasooriya+2022} and briefly reviewed in \S\ref{sec:sam}) that reproduce the properties of observed dwarfs of the Milky Way down to ultra-faint dwarfs. The goal of this work is to test whether these same astrophysical prescriptions and parameters can reproduce the observed cumulative luminosity function and the luminosity--metallicity relation of the Centaurus~A satellites, and, if it can, make predictions for their properties and star formation histories (SFHs), in an effort to investigate the effects of host environment on these dwarfs. \\

The remainder of this paper is organized as follows. In Section \ref{sec:sim}, we describe the details of the N-body simulation of a Centaurus A analog and describe our the semi-analytic model, including the implementation of the extended Press-Schechter (EPS) merger trees, galactic structure, and velocity dispersion. In Section~\ref{sec:obs}, we describe the sample of observational data taken from the literature. Next, in Section~\ref{sec:results}, we compare our models to the observed properties of the known Centaurus~A dwarfs before exploring their potential star formation histories.  Lastly, we present our discussions and conclusions in Section~\ref{sec:conc}.
    
\section{Methods}\label{sec:sim}

\subsection{N-body Simulation}

To carry out of modeling of the Centaurus~A system, we make use of a high-resolution cosmological zoom-in N-body simulation of an isolated Centaurus~A halo from \cite{bovill2016}. To generate this simulation, first a cosmological simulation was run using a 100~$h^{-1}$~Mpc periodic cube with $N = 256^3$ particles, allowing potential Centaurus~A candidate halos to be resolved at $z\,=\,0$ with over $1000$ particles. This simulation was run from $z\,=\,150$ to $z\,=\,0$  with using the WMAP9 cosmology ($\sigma_8=0.821,\,H_0=\,70.0\, \mathrm{km}\,\mathrm{s}^{-1}\,\mathrm{Mpc}^{-1},\,\Omega_\mathrm{b}=0.0463,\,\Omega_\mathrm{m}=0.279,\,\Omega_{\Lambda}=0.721$). Initial conditions were generated using \texttt{MUSIC} \citep{Hahn2011} and the simulation was run with \texttt{Gadget 2} \citep{gadget}. Dark matter halos were identified in the simulation and their properties (including masses, NFW scale radii, spin vectors, positions, and velocities) were calculated using the \texttt{AMIGA} halo finder  \citep{AHF}. These halos were then linked through cosmic time to create merger trees using \texttt{CONSISTENT\_TREES} \citep{consistent_trees}.\\

From this set of halos, a Centaurus~A analog was selected by seeking halos with mass\footnote{As described in the Introduction, the mass of Centaurus~A is poorly constrained observationally. Furthermore, the estimates that have been made \protect\citep{van2000,peng+2004,woodley2007,lokas,Harris+2015,Pearson+2022} have adopted a variety of definitions of virial mass. Given the lack of strong constraints, we choose a value of $10^{13}\mathrm{M}_\odot$, defined as the mass enclosed by a spherical region with density contrast appropriate to the spherical collapse model; \protect\citealt{Bryan&Norman11998}. We will adopt this same definition of virial mass throughout this work.} $\sim~10^{13}~\mathrm{M}_\odot$ that have no halos with mass greater than $10^{12}\;\mathrm{M}_{\odot}$ within $3\;h^{-1}\;\mathrm{Mpc}$ at $z\,=\,0$. The Lagrangian region of this halo (corresponding to approximately 4 times the virial radius (2.4~Mpc) at $z \,=\, 0$) was then resimulated to produce a higher resolution region that covers two times the virial radius at $z\,=\,0$ with no contamination by low resolution particles. This resimulation was performed using an effective number of particles $N_\mathrm{eff} = 8192^3$, corresponding to a particle mass of $m_\mathrm{p} = 1.4 \times 10^5 \mathrm{M}_{\odot}$, using a force softening length of $\epsilon = 200$~$h^{-1}$~pc. This resolution ensures that halos of mass greater than $10^7\,\mathrm{M}_{\odot}$ are resolved with at least 100 particles.

\subsection{Semi-Analytic Model}\label{sec:sam}

%We apply the \textbf{scaled} astrophysical prescriptions and parameters which reproduce the star formation physics and star formation histories of the Milky Way satellites \citep{Weerasooriya+2022} to the N-body merger trees and Extended Press Schechter (EPS) merger trees of Centaurus~A analogs. We make this assumption because there are very few studies exploring the star formation \textbf{physics and } histories of Centaurus~A dwarfs \citep{cote+2009,crnojevic+2011}.\\
 
We model the baryonic content of the Centaurus~A system using the semi-analytic model (SAM) \textsc{Galacticus} \citep{galacticus}. We apply the same astrophysical prescriptions and parameters that  we found were able to accurately reproduce the Milky Way satellites in \cite{Weerasooriya+2022}. That is, we assume that the underlying physics that determines dwarf galaxy formation is the same in both the Milky Way and Centaurus~A systems\footnote{Of course, the prescriptions adopted in \protect\cite{Weerasooriya+2022} all have physically-motivated scalings. For example, the effects of ram pressure stripping scale in proportion to the square of the orbital velocity of each dwarf galaxy, and so will be greater in Centaurus~A than in the Milky Way for any given dwarf.}. \\

Of these astrophysical prescriptions, quenching due to ram pressure, and tidal stripping primarily determine how dwarf satellites are affected by the environment of their host halo. For example, ram pressure stripping can strip gas out of dwarf galaxies resulting in a shortage of gas supply and eventual quenching of star formation. In \cite{Weerasooriya+2022} we found that ram pressure stripping was ineffective at quenching Milky Way satellite galaxies, even when made as efficient as physically plausible. Regardless, in this work we implement ram pressure stripping  with the same high efficiency as in \cite{Weerasooriya+2022}.\\

As discussed in the Introduction, the virial mass of Centaurus~A is highly uncertain. To explore the effects of this uncertainty on our model predictions we wish to investigate a range of possible halo masses for Centaurus~A, consistent with the wide range of observational estimates. As we have only a single N-body realization of the Centaurus~A halo, to explore a range of halo masses we make use of extended Press-Schechter \citep[EPS,][]{Press&Schechter1974,1991MNRAS.248..332B,1991ApJ...379..440B,1993MNRAS.262..627L} merger trees spanning the full range of possible Centaurus~A halo masses ($5\times10^{12}\,\mathrm{M}_{\odot}$ to $1\times10^{13}\,\mathrm{M}_{\odot}$) and constructed using the algorithm of \citeauthor{2008MNRAS.383..557P}~(\citeyear{2008MNRAS.383..557P}; see also \citealt{Cole+2000}) with a minimum resolved halo mass of $1.41\times 10^7\,\mathrm{M}_{\odot}$---corresponding to the lowest mass halos resolved in our high-resolution N-body simulation. Furthermore, we generate 30 realizations at each halo mass, allowing us to explore the halo-to-halo variations in the dwarf galaxy population. While these EPS trees allow us to efficiently probe a range of $M_\mathrm{vir}$ for the Centaurus~A halo, they do not provide detailed positional information on the satellites. As such, EPS trees, unlike their N-body counterparts, do not allow us to explore the dependence of satellite populations on their distance from the host center. From this point onward we will refer to merger trees generated through the Extended Press-Schechter method as ``EPS trees'' and the merger tree from N-body simulation of \cite{bovill2016} as an ``N-body tree''.\\

For a complete description of the \textsc{Galacticus} SAM we refer the reader to \cite{galacticus}. However, in the following subsections we will briefly review key components of the model that are directly relevant to this work. We note that, other than the determination of halo concentrations (see \S\ref{sec:galacticStructure}), all halo and galaxy physics is implemented and applied identically for N-body and EPS merger trees.

\subsubsection{Galactic Structure}\label{sec:galacticStructure}

Each halo in a merger tree is considered as a potential site of galaxy formation (if gas is able to accrete into and cool within the halo---see \citealt{Weerasooriya+2022}). Each galaxy is approximated as consisting of one or both of a rotationally-supported disk and a pressure-supported spheroid. Cooling gas initially settles into a disk, where it may form stars, but can later become part of the spheroid through the action of dynamical instabilities in the disk or mergers with other galaxies.\\

To describe the density distribution of the system we model the dark matter halo as an NFW distribution \citep{1997ApJ...490..493N}, adiabatically compressed by the gravitational potential of the disk and spheroid following the algorithm of \cite{2004ApJ...616...16G}. In the case of N-body merger trees, the NFW scale radii of halos are measured by the \textsc{ConsistentTrees} code and are used directly in our calculations. For EPS trees we instead make use of the concentration--mass-redshift relation of \cite{Diemer&Joyce2019} which was calibrated to accurately match the distribution of halo concentrations measured in N-body simulations. In this model the concentration of a halo is given by:
\begin{equation}
\mathbf{
c=C(\alpha_{\mathrm{eff}})\times\,\tilde{G}\left(\frac{A(n_{\mathrm{eff}})}{\nu} \left[1+\frac{\nu ^2}{B(n_{\mathrm{eff}})}\right]\right),
}
\end{equation}
where
\begin{equation}
\mathbf{
    n_\mathrm{eff}(M) = \left. \frac{\mathrm{d} \log P(k)}{\mathrm{d} \log k}\right|_{k=\kappa 2 \pi/R_\mathrm{L}},
}
\end{equation}
$R_\mathrm{L}$ is the Lagrangian radius of a halo of mass $M$, $P(k)$ is the linear-theory power spectrum as a function of wavenumber $k$, $\nu = \delta_\mathrm{c}/\sigma(M)$ is the peak-height parameter, $\delta_\mathrm{c} \approx 1.686$ is the critical overdensity for collapse of a spherical top-hat perturbation, $\sigma(M)$ is the fractional root variance of the density field in spheres containing, on average, a mass $M$, $A(n_{\mathrm{eff}})=a_0[1+a_1(n_{\mathrm{eff}}+3])$, $B(n_{\mathrm{eff}})=b_0[1+b_1(n_{\mathrm{eff}}+3])$,
$C(\alpha_{\mathrm{eff}})=1-c_{\alpha}(1-\alpha_{\mathrm{eff}})$, and $\tilde{G}$ is the inverse of $G(x)=x/g(x)^{(5+n_{\mathrm{eff}})/6}$, with $g(x) = \log(1+x)-x/(1+x)$. For the constants $\kappa$, $a_0$, $a_1$, $b_0$, and $b_1$ we take the best-fit values found by \cite{Diemer&Joyce2019}.\\

This fitting function gives the concentration of a halo for the case where the virial radius is defined as that radius that encloses a mean density of 200 times the critical density of the universe. Using the NFW profile, we convert this to the corresponding concentration under our preferred definition of the virial radius. \citeauthor{2020Natur.585...39W}~(\citeyear{2020Natur.585...39W}; their Fig.~3) compare the predictions of the \cite{Diemer&Joyce2019} concentration--mass relation to results of N-body simulations down to very low halo masses. For masses above $10^7\mathrm{M}_\odot$ (the lowest halo mass considered in this work) the \cite{Diemer&Joyce2019} fitting function is accurate to better than 0.05~dex.

Galaxy disks are modelled as vertically-thin, exponential disks. The radial scale length of each disk is found by requiring that the disk be rotationally supported in the combined gravitational potential of the dark matter halo, disk and spheroid. To do this we follow the general approach of \cite{Cole+2000}. Briefly, we first compute the specific angular momentum of the disk. Angular momentum is supplied to the disk when gas from the circumgalactic medium (CGM) cools and inflows into the disk---the specific angular momentum of CGM gas is assumed to be comparable to that of the dark matter halo, and to be conserved as the gas flows into the disk. Angular momentum is also lost from the disk due to outflows and transport of disk material into the spheroid. Full details can be found in \cite{galacticus}, \cite{Weerasooriya+2022}, and \cite{2023arXiv230813599A}. Once the specific angular momentum of the disk is known, we solve for a scale radius, $r_\mathrm{d}$ which satisfies:
\begin{equation}
\mathbf{
r_\mathrm{d} \left[ V_\mathrm{d}^2(r_\mathrm{d}) +  V_\mathrm{s}^2(r_\mathrm{d}) +  \tilde{V}_\mathrm{h}^2(r_\mathrm{d}) \right]^{1/2} = j_\mathrm{d},
}
\end{equation}
where $j_\mathrm{d}$ is the specific angular momentum of the disk at its scale radius (assuming a flat rotation curve), and $V_\mathrm{d}(r)$, $V_\mathrm{s}(r)$, and $\tilde{V}_\mathrm{h}(r)$ are the rotation curves due to the disk, spheroid, and adiabatically-contracted halo respectively.

Spheroids are modeled as \cite{1990ApJ...356..359H} profiles. The scale radius of this profile is found using a similar approach as for the disk. Even though the spheroid is pressure-supported, we compute a pseudo-specific angular momentum---essentially the specific angular momentum that the spheroid would have if it \emph{were} rotationally supported. We then solve for the scale radius in the same way as described above for the disk, but using this pseudo-specific angular momentum. Full details of this approach are given in \cite{Cole+2000}.
As these rotational support calculations must take into account the self-gravity of disk and spheroid components (and because of the inclusion of adiabatic contraction of the dark matter halo), the above equations are solved using an iterative method at each evolution step in our SAM.

\subsubsection{Line-of-sight Velocity Dispersion}\label{sec:velocityDispersion}

Given the structure of a halo+galaxy system as computed in \S\ref{sec:galacticStructure} we can estimate a line-of-sight-velocity dispersion for the spheroid component. To do so we assume that the spheroid is in Jeans equilibrium (with an isotropic velocity distribution) in the combined gravitational potential of the adiabatically-contracted halo, disk and its own self-gravity. For dwarf galaxies the gravitational potential is largely dominated by the contribution of their dark matter subhalo. \cite{1996MNRAS.281..716C} have shown that Jeans equilibrium provides an accurate description of the velocity dispersion as a function of radius measured for N-body dark matter halos.

We first solve for the 1D velocity dispersion as a function of radius, $\sigma(r)$, in the spheroid by integrating the Jeans equation from infinity to each radius. Then, to estimate an observable velocity dispersion, we compute the density-weighted line-of-sight velocity dispersion at the half-mass radius, $r_{1/2}$, of the Hernquist profile for each spheroid:
\begin{equation}
\sigma^2_\mathrm{los} = \left. \int_{r_{1/2}}^\infty \sigma^2(r) \rho(r) {r \over \sqrt{r^2-r_{1/2}^2}} \mathrm{d}r \right/ \int_{r_{1/2}}^\infty \rho(r) {r \over \sqrt{r^2-r_{1/2}^2}} \mathrm{d}r,
\end{equation}
where $\rho(r)$ is the Hernquist profile of the spheroid.

\subsubsection{Spectra and Magnitudes}\label{sec:spectraMagnitudes}
 
\cite{Weerasooriya+2022} describes in detail how we model star formation in our model. Once the star formation rate as a function of time, $\phi(t)$, in a galaxy disk or spheroid is known we compute the stellar spectrum of that component as:
\begin{eqnarray}
F(\lambda) = \int_0^t \phi(t^\prime) S(\lambda | t-t^\prime, Z[t]) \mathrm{d}t^\prime,
\end{eqnarray}
where $Z[t]$ is the metallicity of stars forming at time $t$, and $S(\lambda | \tau, Z)$ is the spectrum of a unit mass, single age, single metallicity stellar population of age $\tau$ and metallicity $Z$. For $S(\lambda | \tau, Z)$ we adopt the models of \citeauthor{2009ApJ...699..486C}~(\citeyear{2009ApJ...699..486C}; as implemented by \citealt{2010ascl.soft10043C}) assuming a \cite{2001ApJ...554.1274C} initial mass function. 

Observed luminosities in any band of interest\footnote{In this work we will consider the standard V-band, along with the SDSS g-band, and DES g-band.} are then found by integrating this spectrum under the appropriate filter response curve. This allows us to find the luminosity of the disk and spheroid component of each galaxy in a band of interest. Given these, the structural properties of disk and spheroid (found as described in \S\ref{sec:galacticStructure}), and the assumption that the starlight in each component follows the same radial distribution as the mass (i.e. an exponential disk and a \cite{1990ApJ...356..359H} spheroid) we can numerically solve for the half-light radius of each galaxy in any band of interest.

\section{Observational Sample} \label{sec:obs}

Observations of the Centaurus~A dwarfs are inhomogeneous, therefore in this section we describe the sample of Centaurus~A dwarfs taken from the literature that we will compare to our models. We use observational data for Centaurus~A satellites from a variety of sources including \cite{crnojevic2010,crnojevic2014,crnojevic2016,crnojevic2019,karachentsev2013,muller2015,muller2017,Muller2019,SCABS,Taylor+2018}, which are presented in Table \ref{tab:table}. In addition to the dwarfs in the above sample, we also include 38 new dwarf galaxy candidates from \cite{Taylor+2024} in our overall analysis, but do not list their properties in Table~\ref{tab:table}. This new sample includes only those galaxies with distance or velocity measurements that verify them as members of Centaurus~A with distances $\leq 5.8\,\mathrm{Mpc}$.\\

\begin{figure*}[h!]
    \centering
    \includegraphics[width=15cm]{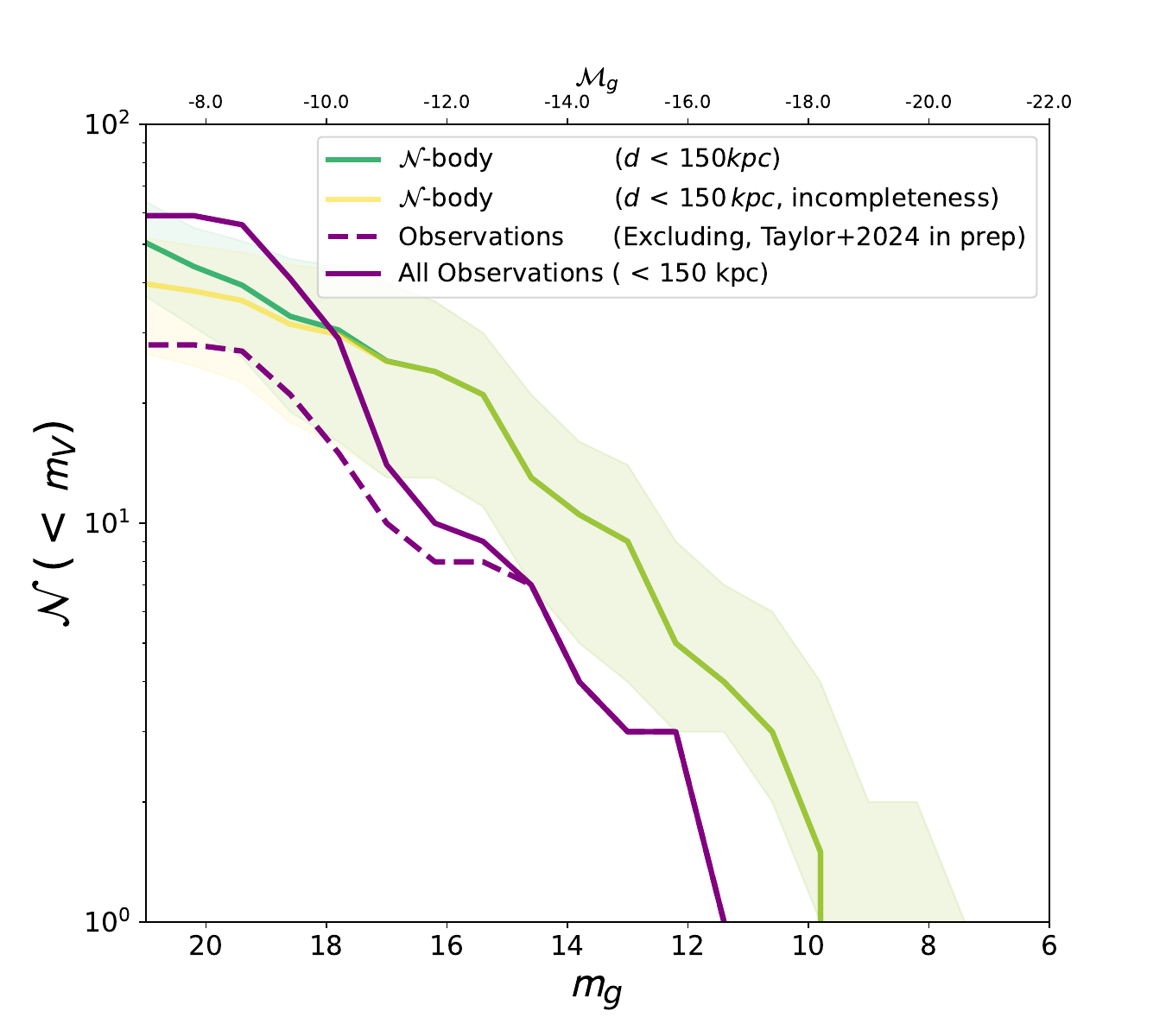}
    \caption{The cumulative, V-band luminosity function as a function of apparent magnitude within a projected distance of 150~kpc of Centaurus~A, corresponding to the region which is well-covered by the SCABS and PISCeS surveys. The $y$-axis shows the number of galaxies brighter than a given V-band apparent magnitude. For reference, the corresponding absolute magnitude at the distance of Centaurus~A is shown on the upper axis. The solid purple curve shows all known observed galaxies, including those from SCABS (including the SCABS dwarfs from \citealt{Taylor+2024}) and PISCeS.  The dashed purple curve excludes new dwarfs from the sample of \cite{Taylor+2024}. Predictions from our model, using our N-body merger tree, are indicated by the green curve and shaded regions, which represent the median and full extent over all viewing angles. The yellow curve shows the same model results, but accounting for the observational completeness fractions as calculated by Leahy et al. (private communication).} 
    \label{fig:completeness}
\end{figure*}

Absolute magnitudes of observed Centaurus~A dwarfs are reported in a variety of different bands. To facilitate some comparisons with our model, we convert these observations to a common band, for which we choose the V-band. To perform this conversion, consider an observed galaxy with a reported absolute magnitude in the ``X'' band (where ``X'' here can be any observed band) of $M_\mathrm{X}$. We compute the median V$-$X color of Centaurus~A dwarfs from observations, $\overline{M_\mathrm{V}-M_\mathrm{X}}$, including all Centaurus~A dwarfs with absolute magnitudes measured in both the V and X bands. We then approximate the V-band absolute magnitude of the observed dwarf as $M_\mathrm{V} \approx M_\mathrm{X} + \overline{M_\mathrm{V}-M_\mathrm{X}}$. Any galaxy lacking an observed V band magnitude, but with observed magnitude in some other band X are converted to V-band magnitudes in this way. We apply this conversion only to dwarfs lacking V-band magnitude data in the literature, and only for Figures \ref{fig:completeness} and \ref{fig:lum_600}. In other figures, we show only those dwarfs for which measured V-band magnitudes \emph{are} available  in the literature.\\

The current sample of observations of Centaurus~A is incomplete due to three major reasons: 1.) Lack of systematic sky coverage that leads to spatial incompleteness. For example the PISCeS survey is spatially incomplete, and biased toward coverage of the Northeastern region of Centaurus~A's halo. These non-uniformities in design or analysis make completeness calculations more complex. 2.) The detectability limits of different surveys, e.g. SCABS is limited to $M_\mathrm{V}<-7.2$ within $150$~kpc covering an area of 50 square degrees and surface brightness limit of $27.8\, \mathrm{mag\,arcsec}^{-2}$ in the g band (Leahy et al., private communication), while PISCeS can detect dwarfs down to $M_\mathrm{V}<-8$ within $150$~kpc and with a surface brightness limit of $26.5\, \mathrm{mag\,arcsec}^{-2}$ in the g-band \citep{crnojevic2014,crnojevic2019}. 3.) The lack of distance measurements in the outskirts also hinders the determination of membership \citep{muller2017,Taylor+2018}. Distances are essential to determine the membership of satellites, their shape, and their brightness. Thus larger uncertainties in  distances can cause these data to have larger uncertainties in their sizes and magnitude measurements. However, the observational sample is relatively complete within 200 kpc \citep{Muller2019} down to $M_\mathrm{V}\,=\,-10$. The SCABS and PISCeS surveys cover a spatial region within a projected radius of 150~kpc down to dwarfs as faint as $M_\mathrm{V}<-7.2$ \citep{crnojevic2014,crnojevic2016,SCABS}.\\

Under the definition adopted in this work, and assuming a mass of $10^{13}\mathrm{M}_\odot$ for the halo of Centaurus~A, the corresponding virial radius is 562~kpc (somewhat larger than the estimate of $\sim 409$ kpc from \cite{Tully2015} but comparable given the large uncertainties in the halo mass). However, it is expected that many satellites of Centaurus~A will be found beyond this radius, yet within Centaurus~A's `splashback radius'. The splashback radius is a physically motivated halo boundary that eliminates spurious evolution of radius and mass caused by standard definitions of virial radius \citep{Diemer+2014,Deason+2020,Shin&Diemer2023}. We estimate the splashback radius of Centaurus~A using the model of \cite{2020ApJS..251...17D}, which predicts\footnote{\protect\cite{2020ApJS..251...17D} show that the splashback radius depends on a halo's accretion rate. Lacking any knowledge of the accretion rate of Centaurus~A's halo we adopt the mean value for halos of its mass, also computed using the model of \protect\cite{2020ApJS..251...17D}.} a splashback radius of 712~kpc. Therefore we will also consider model predictions within a projected radius of 700~kpc to approximately capture the entirety of the satellite population within this splashback radius. While current observations are highly incomplete in this region---limiting any comparison between model and observations---our results serve as testable predictions for future observational campaigns.\\
    
The observational properties of Centaurus~A satellites are still largely unknown. Thus our knowledge of these galaxies and their properties is nowhere near as complete as our knowledge of the Milky Way system. Quantitative estimates of the incompleteness of Centaurus~A dwarfs have not been made by previous studies, and such exploration is beyond the scope of our current work. A preliminary quantitative exploration of the completeness limits in SCABS Centaurus~A satellites is underway by Leahy et al. (private communication). These completeness tests are preliminary results based on artificial galaxy experiments using an automated dwarf galaxy selection technique. These results are based on an analysis in the g'-band and are applied to a small subset of the overall SCABS imaging that does not extend beyond 150 kpc. Their exploration of completeness using 5000 Monte Carlo dwarf galaxy realizations reveals completeness of $96\%$ for dwarf galaxies of $\lesssim18\,\mathrm{mag}$ in the g-band. They report $50\%$ completeness at g-band magnitudes of 20.01 and $50\%$ completeness at surface brightness of $\sim27.8\, \mathrm{mag\,arcsec}^{-2}$. In Figure \ref{fig:completeness}, we compare our models within 150~kpc to the total number of galaxies expected in the region based on these preliminary artificial galaxy experiments. While this region only covers a fraction of the entire virial volume of Centaurus~A, it provides an almost complete observational census of dwarfs with which to compare our model. Note that, as we must select model galaxies based upon their projected position relative to the center of their host halo, we show results only for our model applied to the N-body simulation, for which such positions are available. The green curve shows the number of satellites predicted within 150 kpc of Centaurus~A, and the yellow curve shows the same model with the incompletenesses of Leahy et al. (private communication) applied to that model. This curve signifies the number of satellites predicted by the model within 150~kpc that would be observable. As there is no preferred viewing angle of our modeled Centaurus~A, we rotate the model distribution of galaxies around the $\hat{z}$-axis of the simulation in 5-degree steps, project the galaxy positions into the $y$--$z$ plane, and repeat the projected radius selection at each step. This provides an estimate of the uncertainty in the predicted luminosity function due to the anisotropy of the system. This error envelope is indicated by the shaded region. The purple curves show the observed dwarfs within 150 kpc of Centaurus~A, with the solid line showing all observations and the dashed line excluding the dwarfs of \cite{Taylor+2024}. We note here that given the unconfirmed natures of the \cite{Taylor+2024} dwarf candidates, this estimate should be considered as an upper limit with results interpreted in that context. Based on the comparison, our galaxy models agree with the observed luminosity function, although the observed luminosity function demonstrates a slightly steeper slope within this 150~kpc radius. Nevertheless, the lower boundary of the luminosity function envelope that accounts for different projections along the line of sight generated from the N-body model is consistent with observations for the fainter dwarfs (see also Figure~\ref{fig:Mv_200}). At brighter magnitudes ($M_\mathrm{V} < -14$), our model predicts marginally more galaxies than are observed. This is consistent with other studies of the inner 200 kpc of the Centaurus~A halo \citep{Muller2019}. 

We reiterate that the observational knowledge of the dwarf galaxy population of Centaurus~A beyond 150~kpc remains highly incomplete and the completeness of the galaxy sample in these outer regions is not well-characterized\footnote{The galaxies beyond the regions surveyed by SCABS and PISCeS are drawn from a variety of different literature sources. Completeness estimates are not available for any of these samples and, in most cases, even the area of sky surveyed is unclear, making any correction for incompleteness and survey area impossible.} In much of the remainder of this work we will present model predictions throughout the entirety of the Centaurus~A system (out to the splashback radius). These should be viewed as testable predictions for future observational campaigns.

\begin{table*}[h!]
    \centering
    \begin{tabular}{cccccccccccc}
    \hline
    Name & RA$_\mathrm{h}$ & RA$_\mathrm{m}$ & RA$_\mathrm{s}$ & Dec$_\mathrm{deg}$ & Dec$_\mathrm{arcmin}$ & Dec$_\mathrm{s}$ & $M_\mathrm{B}$ & $M_\mathrm{r}$ & $M_\mathrm{V}$ & Distance & References \\
     & (h) & (min) & (s) & (deg) & (arcmin) &(arcsec) & (mag) & (mag) & (mag) & (Mpc) & \\
    \hline
    NGC4945  & 13 & 5 & 26.1 & -49 & 28 & 16.0 & -20.34 & -18.2 & -20.6 & 3.47 & 1,7 \\
    NGC5102 & 13 & 21 & 57.8 & -36 & 37 & 47.0 & -18.24 & -20.13 & -20.37 & 3.66 & 1,7 \\
    E383-087 & 13 & 49 & 17.5 & -36 & 3 & 48.4 & -16.83 & -17.17 & -17.31 & 3.19 & 1,7 \\
    NGC5206 & 13 & 30 & 41.0 & -47 & 53 & 42.0 & -16.43 & -16.97 & -16.90 & 3.6 & 1,7 \\
    NGC5408 & 14 & 0 & 18.0 & -41 & 8 & 11.0 & -15.91 & -16.27 & -17.3 & 4.81 & 2,7 \\
    E324-24 & 13 & 27 & 37.4 & -41 & 28 & 50.0 & -15.49 & -15.41 & -15.6 & 3.78 & 2,7 \\
    E26958  & 13 & 7 & 38.0 & -46 & 43 & 30.0 & -14.99 & -17.09 & -16.8 & 3.63 & 6,7 \\
    NGC5237 & 13 & 37 & 38.9 & -42 & 50 & 51.0 & -14.82 & -15.27 & -15.08 & 3.33 & 3,5,7 \\
    NGC5011C  & 13 & 13 & 11.9 & -43 & 15 & 56.0 & -14.15 & -14.19 & -15.8 & 3.73 & 2,3,7 \\
    E325-11 & 13 & 45 & 0.8 & -41 & 51 & 32.0 & -14.02 & -14.47 & -14.5 & 3.4 & 3,6,7 \\
\hline
    \end{tabular}
    \caption{Properties of galaxies in the Centaurus A system. Column 1 gives the name of each galaxy, columns 2--4 show right ascension, columns 5--7 declination, columns 8--10 B-, r-, and V-band absolute magnitude respectively, column 11 gives distance, and column 12 lists the references from which this information was drawn. References: (1) \cite{Lauberts+1989}, (2) \cite{deVacouleurs+1991},(3) \cite{Karachentsev+2003},(4)
\cite{James+2004}, (5) \cite{Doyle+2005}, (6) \cite{Sharina+2008}, (7) \cite{karachentsev2013}, (8) \cite{muller2015}, (9) \cite{muller2017}, (10) \cite{crnojevic2014,crnojevic2016,crnojevic2019}, (11) \cite{Taylor+2018}.
Table 1 is published in machine-readable format. Only a portion of this table is shown here to demonstrate its form and content.}

    \label{tab:table}
\end{table*}

\section{Properties of Centaurus~A dwarfs} \label{sec:results}
% While the observations above are incomparable to that of the MW satellites, we attempt to model the Centaurus~A dwarf galaxy population as best we can. 
In this section, we compare different properties of the modeled Centaurus~A dwarfs with observational data described in Section \ref{sec:obs}. We start by exploring the properties of the Centaurus~A dwarfs including luminosities, half-light radii, and velocity dispersions.

\subsection{Luminosity Function}

\begin{figure*}[h!]
    \centering
    \includegraphics[clip=true,trim={0cm 0cm 0 0cm},width=\textwidth]{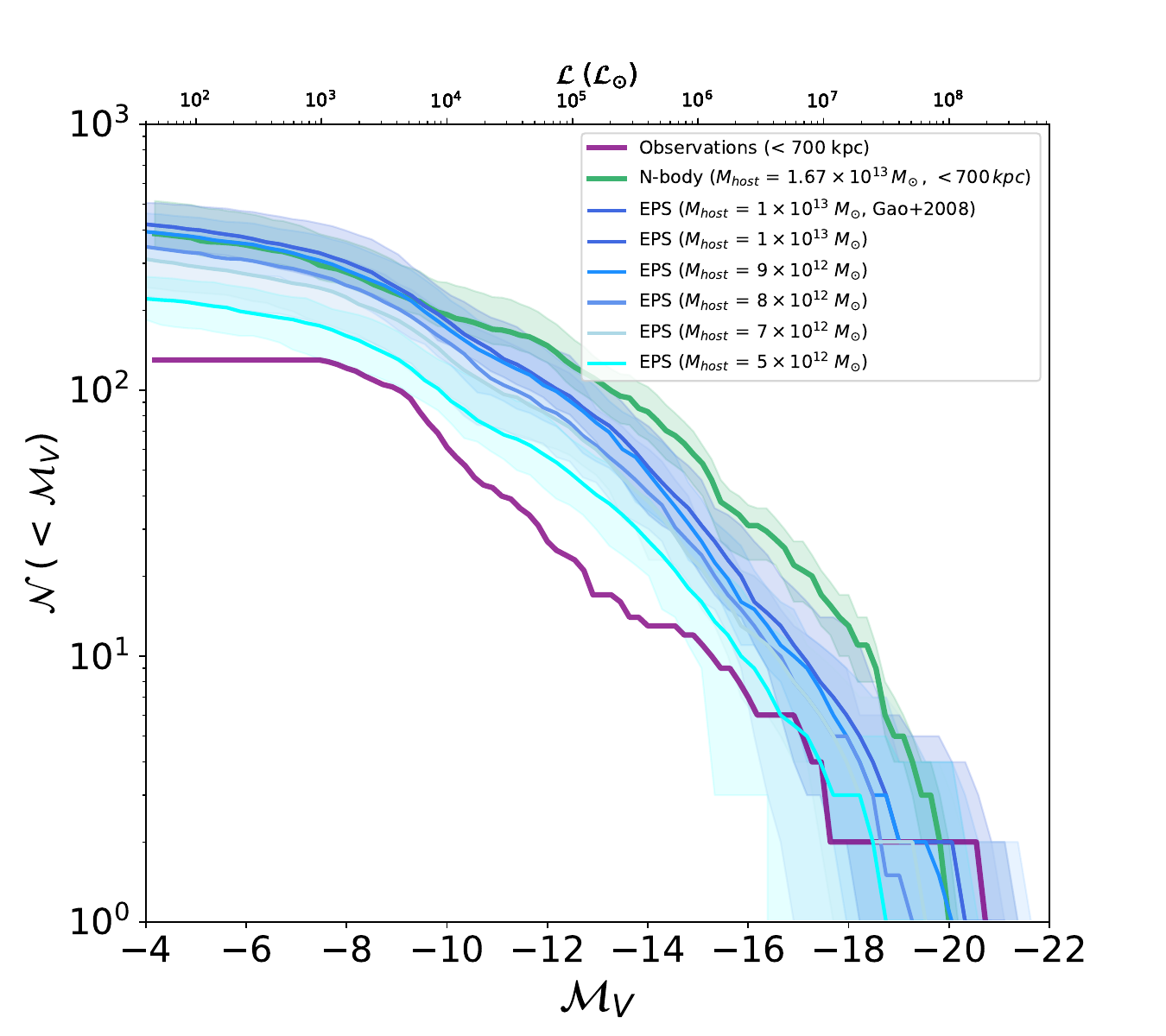}
    \caption{Luminosity functions for the Centaurus~A satellites within $\mathbf{700}$ kpc projected radius (around 1.5 times the virial radius, and corresponding to the likely splashback radius of Centaurus~A) for models and observations. The purple line shows the observational sample. The predicted luminosity functions for Centaurus~A satellites with $M_\mathrm{V}\leq -4$ from our N-body merger tree are shown in green, while lines shown in shades of blue represent EPS trees with different Centaurus~A masses as indicated in the figure. Each blue line shows the median luminosity function over 30 different EPS trees, with the shaded region indicating the minimum and maximum.}
    \label{fig:lum_600}
\end{figure*}

In Figure \ref{fig:lum_600}, we plot the cumulative luminosity function of Centaurus~A satellites for observations within a projected radius of $700$ kpc---approximately 1.5 times the virial radius and consistent with the likely splashback radius of Centaurus~A. Selection of model galaxies based on projected radius is possible only when using N-body trees, for which positional information is available. The median luminosity function over all potential viewing angles is shown in green, while the maximum and minimum are shown by the shaded envelope. All observations are converted to V-band magnitudes as described in \S\ref{sec:obs}.\\

We also run models using merger trees generated via an Extended Press-Schechter (EPS) approach to explore the mass range of Centaurus~A's halo. EPS models for different mass merger trees---spanning the plausible range of masses for the Centaurus~A halo---are shown in shades of blue. These models inherently do not have positional information. However, by construction, all galaxies in the EPS trees are within the splashback radius. Therefore we include all the galaxies in merger trees for these models. The solid curves show the median per $M_\mathrm{V}$ bin with shaded areas indicating the minimum and maximum over 30 EPS trees. All of these models follow the general shape of the luminosity function for observations within 700~kpc. However, there is some variation in the shape of the luminosity functions between N-body and EPS merger trees.\\

The purple line in Figure~\ref{fig:lum_600} shows the cumulative luminosity function of known, observed satellite galaxies within 700~kpc of Centaurus~A. As discussed in \S\ref{sec:obs} the observed sample of galaxies in this region is likely to be highly incomplete---more so at faint magnitudes---and no good estimates of completeness (or the fraction of the projected area of the Centaurus~A system surveyed to a given depth) are available. As such, the purple line in Figure~\ref{fig:lum_600} should be considered to be a lower limit to the true luminosity function, particularly at fainter magnitudes.\\

While we cannot constrain the mass of Centaurus~A from the models due to poorly understood completeness in the observations for fainter dwarfs, comparison with models at the bright end of the luminosity function suggests that Centaurus~A is likely to have a higher mass ($10^{13}\,\mathrm{M}_{\odot}$) . Our model is consistent with the overall shape of the observed cumulative luminosity function for both N-body and EPS merger trees at all modeled host masses.\\

\begin{figure*}[h!]
    \centering
    \includegraphics[clip=true,trim={0cm 0 0cm 0 0},width=10cm]{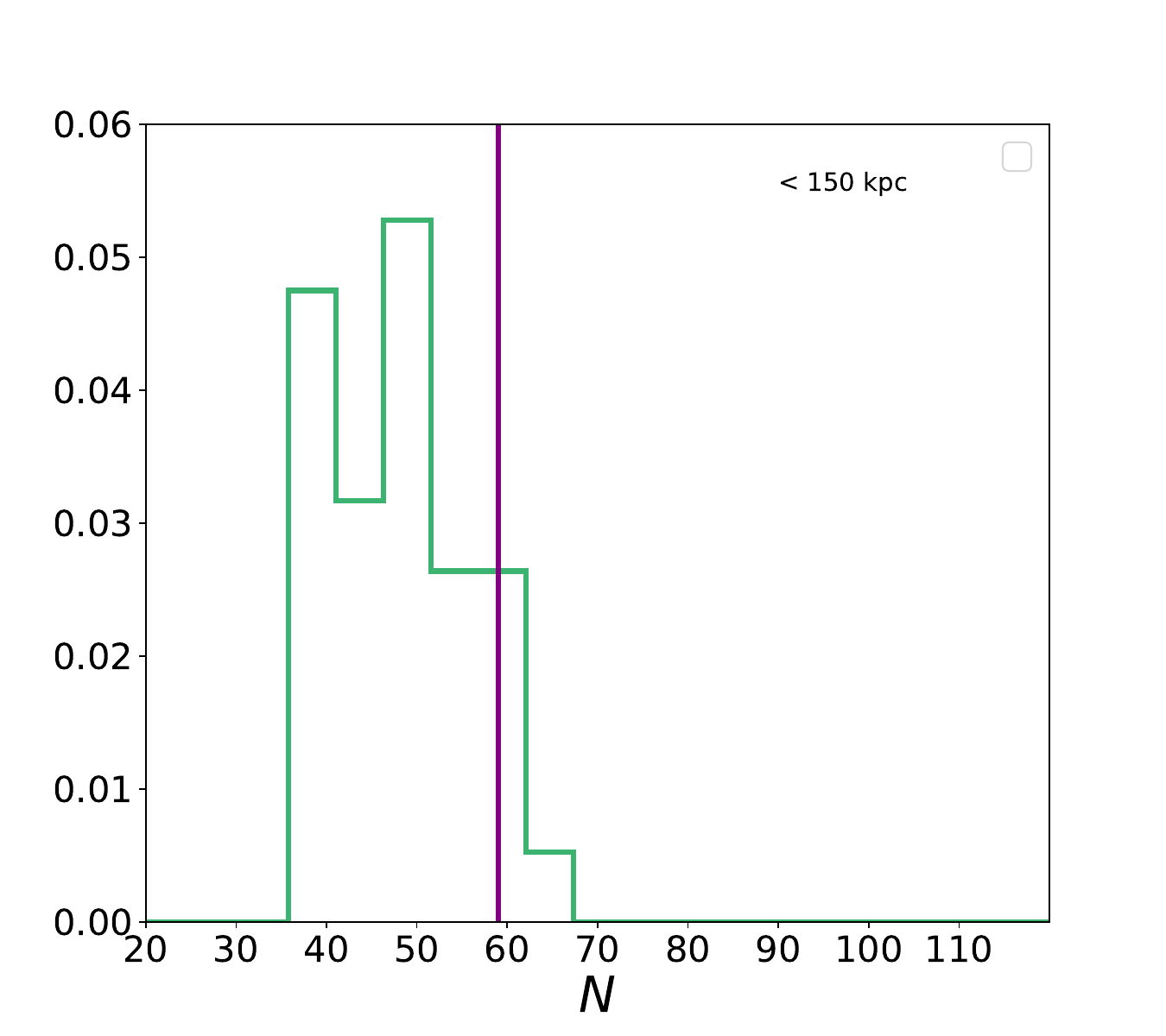}
     \caption{The probability distribution of the number of satellites within a projected distance of 150~kpc of Centaurus~A and brighter than $M_\mathrm{V}\leq -6$ for different lines of sight as the model is rotated in the $y$-$z$ plane (green histogram). The mean and median of this distribution are 48.1 and 49 respectively. The purple line shows the total number of satellites in the observed sample (59).}
     % Right panel: Cumulative luminosity function for the Centaurus~A satellites within a projected distance of 150 kpc. The green line and shaded region show the median and full extent of predictions from our model applied to our N-body merger tree over all viewing angles. The purple line shows the luminosity function of all observed galaxies within a projected distance of 150 kpc.}
    \label{fig:Mv_200}
\end{figure*}

The poorly understood completeness limits of the observations within 700 kpc of Centaurus~A make exploration of these satellites and their properties limited (see section \ref{sec:obs} for a detailed discussion of completeness). In Figure~\ref{fig:Mv_200} we show the distribution of the number of bright ($M_\mathrm{V} \leq -6$) satellites within a projected distance of 150~kpc of Centaurus~A (for which observations \emph{are} largely complete to this magnitude) in our model as the line of sight is rotated relative to the $\hat{z}$ direction in 5 degree increments as the green histogram. For comparison we show the number of observed satellites of the same brightness in this region as a vertical purple line. Our model predicts a median of 49 that is lower than the observed number of dwarf galaxies (59) located within 150~kpc of Centaurus~A, but the observed number is well within the distribution predicted by our model. Perhaps surprisingly, the brightest observed satellites of the Centaurus~A system (which extend to magnitudes of around $M_\mathrm{V}=-21$ as shown in Figure~\ref{fig:lum_600}) are not found within 150 kpc of Centaurus~A. As such, they do not appear in Figure~\ref{fig:completeness} which includes only this inner region. While these very luminous galaxies are not considered dwarfs, and so their spatial distribution is not directly relevant to the focus of this paper, we note that our N-body simulation never predicts zero such galaxies in this region\footnote{Of course, our N-body simulation is just a single realization and so we can not draw strong conclusions from this apparent discrepancy with the observations.}. 
    %
    
    % 
    
    % \item SCABS sample can detect dwarfs $MV<-7.2$ \citep{SCABS}, while PISCeS detect dwarfs $M_\mathrm{V}<-8$ \cite{crnojevic2014}.
    
    % \item Note that the shape of the luminosity function within 150 kpc do not follow observations (second panel of Figure \ref{fig:hist}). That is, there is a dearth of observed luminous satellites brighter than -13 mag. Yet, the number of luminous dwarfs agree with the number of observed galaxies within 150 kpc (first panel of Figure \ref{fig:hist}).
    
%     \begin{figure*}[h!]
%     \centering
%     \includegraphics[clip=true,trim={2cm 0cm 0cm 0cm},width=18cm]{histogram_150kpc.pdf}
%     %left, down, right, up
%     \caption{Left panel: Number of luminous satellites brighter than $M_\mathrm{V}\leq -6$ predicted within 150 kpc for different lines of sight as x-y plane is rotated (shown in green). Purple line shows the total number of satellites in the observed sample. Right panel: cumulative luminosity functions for the Centaurus~A satellites within 150 kpc in the y-z plane are shown in green. We plot the median of the cumulative luminosity in green and shade within our uncertainty budget. }
%     \label{fig:hist}
% \end{figure*}

\subsection{Half Light Radii}

\begin{figure*}[h!]
     \centering
    \includegraphics[clip=true,trim={0cm 0cm 3cm 0cm},width=.48\textwidth]{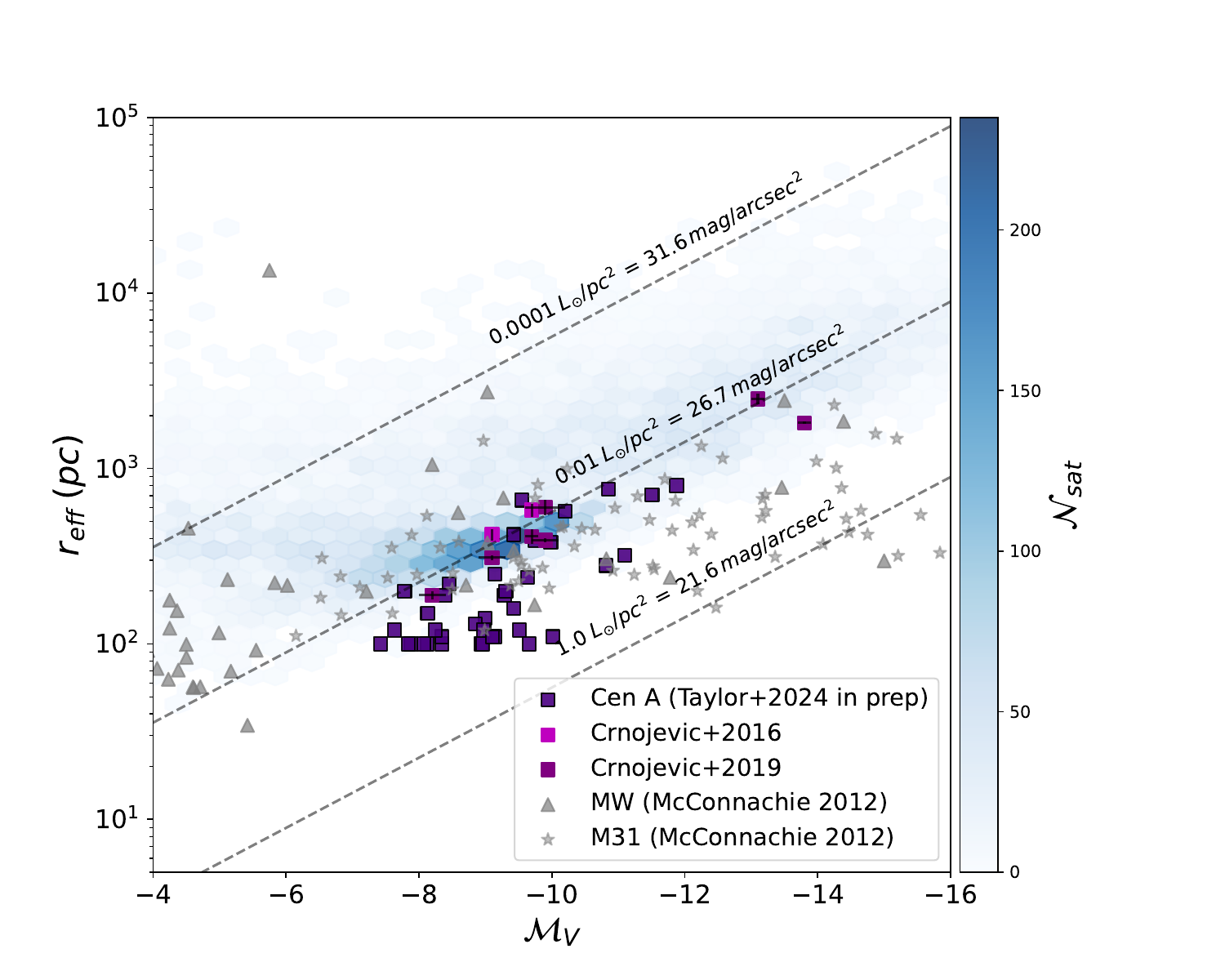}
      \centering
    \includegraphics[clip=true,trim={0cm 0cm 3cm 0cm},width=.48\textwidth]{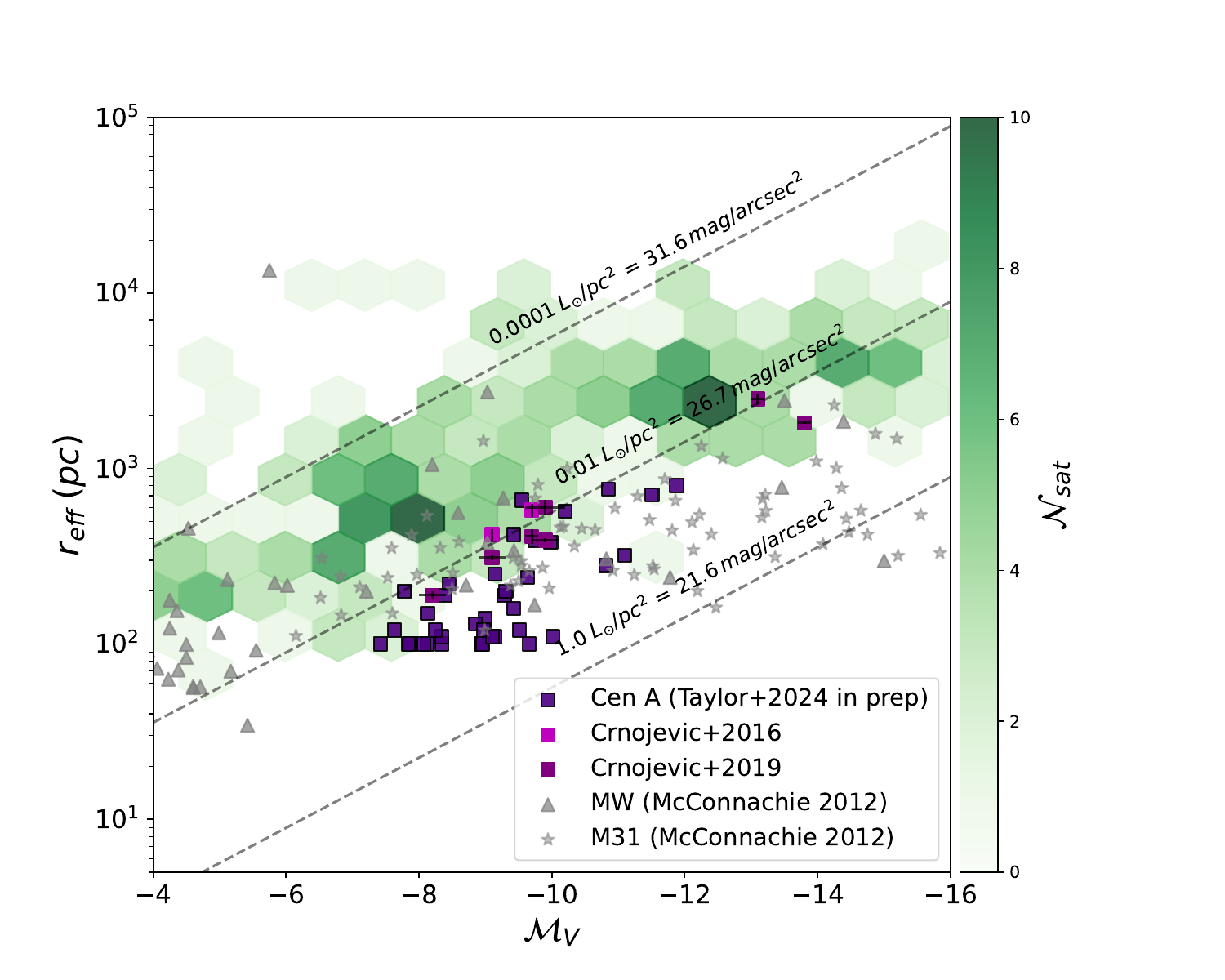}
\caption{Half light radii vs. the absolute V-band magnitude of the Centaurus~A satellites are shown for both modeled and observed galaxies. The left panel shows the distribution of model results using EPS merger trees (blue hexagons), while the right panel shows the distribution of model results using the N-body merger tree within a projected distance of 700 kpc in order to encompass all satellites within the backsplash volume (green hexagons). The contrast of each hexagon indicates the number of model galaxies falling within that region, as indicated by the color bar to the right of each panel. For model galaxies, half-light radii are measured in the $g_{DES}$ band, although the dependence of half-light radius on band is quite small. The same observational data is repeated in each panel. Specifically, we show observed half-light radii of Centaurus~A dwarfs from \cite[measured in the $g_{DES}$-band][]{Taylor+2024} as dark purple squares, and \cite{crnojevic2016,crnojevic2019} (V-band) as magenta squares. The observations of \cite{mcconnachie2012} for the Milky Way (measured in the V-band), and M31 (measured in the V-band) are also shown as grey triangles and stars respectively. Iso-surface brightness lines are shown as grey dashed lines. The predicted trend for half-light radii is consistent with that which our model predicted for the Milky Way satellites.\citep{Weerasooriya+2022}.}
\label{fig:HLR_Mv}
\end{figure*}

%\textbf{Equilibrium radii for disk and spheroid components are computed by finding the radius at which the galaxy is rotationally supported against the combined gravitational potential of itself and the dark matter halo, i.e. $j=GM_{DM}(r)r$ where j is the angular momentum.
%First, the mass of each galaxy is calculated by integrating the NFW density profile of the galaxy from 0 to R. This is then used to resolve for half radius or the radius at which half of the mass is enclosed. It's important to note that the methods for determining scale radii differ between N-body and EPS trees. For N-body-based merger trees, the scale radii are directly obtained from the simulation data, and the concentrations are then calculated. On the other hand, for EPS merger trees, concentrations are determined using a fitting function described by the equations:
%  $\mathrm{log}c=A\mathrm{log}_{10}\mathrm{M}_{halo}+B$ from \cite{Gao+2008}, where 'A' and 'B' are functions of the expansion factor 'a.' It's worth mentioning that this fitting function, as outlined by \cite{Gao+2008}, has been calibrated and validated using N-body simulation data.}

%$$A=-014exp\bigg[-\bigg(\frac{\mathrm{log}_{10}a+0.05}{0.35}\bigg)^2\bigg]$$,
%$$B=2.646exp\bigg[-\bigg(\frac{\mathrm{log}_{10}a}{0.35}\bigg)^2\bigg]$$
%\textbf{Finally, the half-light radii is extracted by weighting the half-mass of the baryonic plus dark matter mass by the corresponding luminosity in the SDSS g band.
%}

In Figure \ref{fig:HLR_Mv}, we compare the half-light radii vs. luminosity for the modeled dwarfs (using the N-body merger tree in the right panel and 30 EPS merger trees with $\sim 1\times 10^{13}\,\mathrm{M}_\odot$ in the left panel) with the observed Centaurus~A dwarfs \citep{Taylor+2024, crnojevic2016,crnojevic2019} along with observations of other systems such as the Milky Way, and M31 \citep{mcconnachie2012} for comparison. We note that the majority of observations beyond the Local Group are not sensitive to surface brightnesses below $0.01\;\mathrm{L}_{\odot}\,\mathrm{pc}^{-2}$.\\

% Galaxy sizes are computed by finding the radius at which  the galaxy is rotationally supported against the combined gravitational potential of itself and the dark matter halo, given the computed angular momentum content of the galaxy. The half-light radii of the modeled Centaurus~A satellite galaxies are computed in the $g_{SDSS}$ band. \textsc{Galacticus} determines the half-light radii using the dark matter profile of halos. Dark matter profile is determined from a NFW profile. The scale radii for trees based on simulations are set from the N-body simulations while for trees based on EPS trees, the concentrations are calculated using the model by \citep{Gao+2008}. We use SDSS filters since SCABS data are calibrated to the SDSS photometric system \citep{SCABS}.\\

At high luminosities, the modeled sizes of the dwarfs agree with observations of the Centaurus~A dwarfs. However, modeled sizes for the fainter dwarfs are larger than the observed values. This is consistent with the systematically larger sizes of the fainter modeled dwarfs compared to observations of the Milky Way satellites \citep{Weerasooriya+2022}. \textsc{Galacticus} tends to overpredict the half-light radii with merger trees generated from EPS and N-body simulations. EPS trees lose accuracy for halos with $\lesssim 10^{10} M_\odot$ due to dynamic range limitations \citep{Sommerville1999,Zhang+2008}. This may lead to inaccuracies in halo masses and/or formation times, which may subsequently affect the sizes of the galaxies forming in these halos. N-body trees over estimate half-light radii if a halo has $N<1000$ particles \citep{Weerasooriya+2022}. A more detailed investigation of how the resolution affects the size of the modeled dwarfs will be a subject of future study.
    % \item In addition, the scatter in the observed half light radii vs $M_\mathrm{V}$ for Centaurus~A system also seems to be high for dwarfs fainter than $M_\mathrm{V}>-10$. Thus more followup observations are required to understand the sizes of Centaurus~A satellites.

    % \item However, they are consistent with several observed Local Group dwarfs of \cite{mcconnachie2012} (see grey dots and blue stars in Figure \ref{fig:HLR_Mv}). Same comparison have been observed by \cite[Figure 3][]{Taylor+2018}. This suggests similarities between the dwarf populations in these systems.

\subsection{Metallicity}

    \begin{figure*}[!h]
         \centering
        \includegraphics[clip=true,trim={0cm 0 1cm 2cm},width=.48\textwidth]{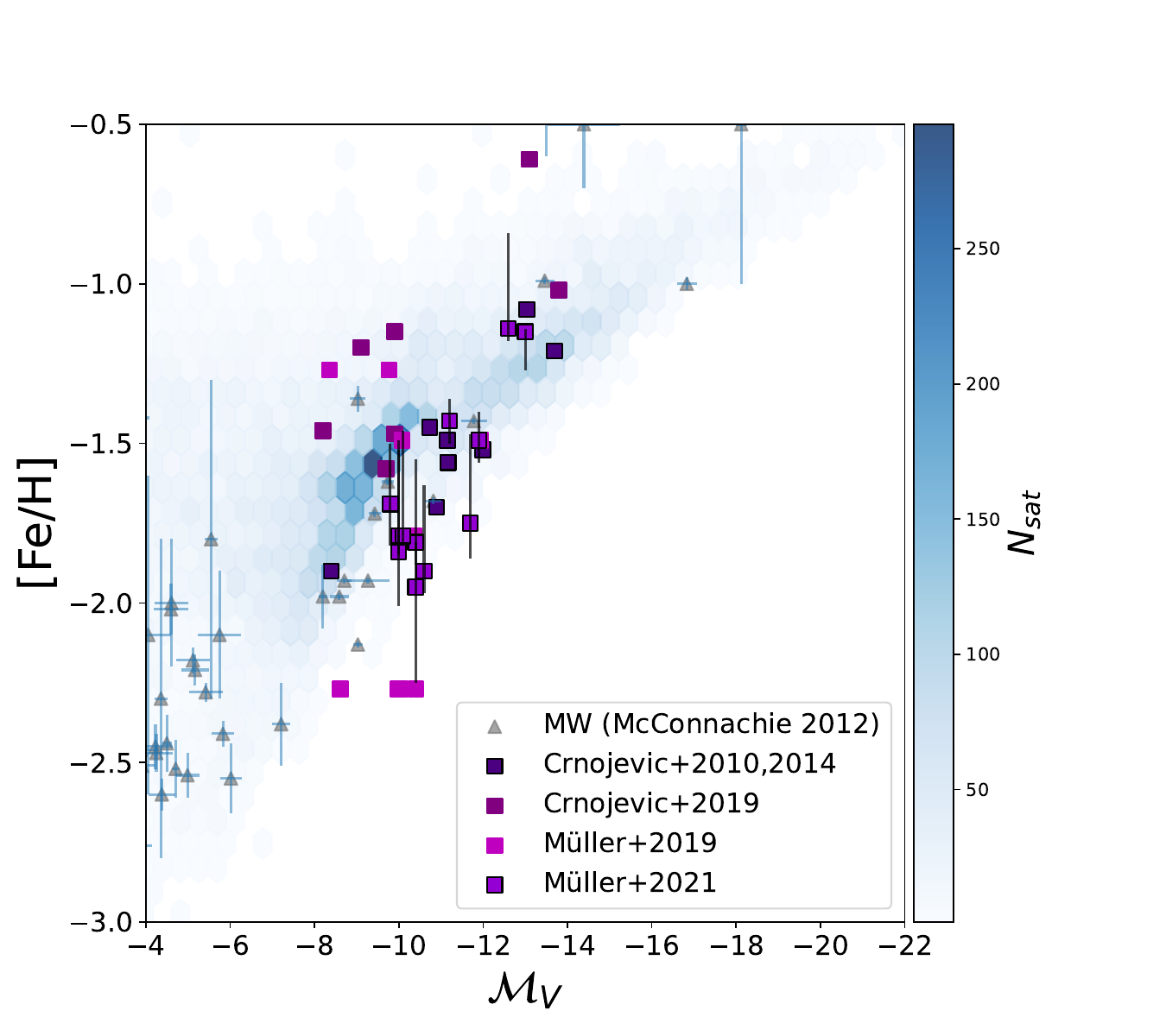}
        \centering
        \includegraphics[clip=true,trim={0cm 0 1cm 2cm},width=.48\textwidth]{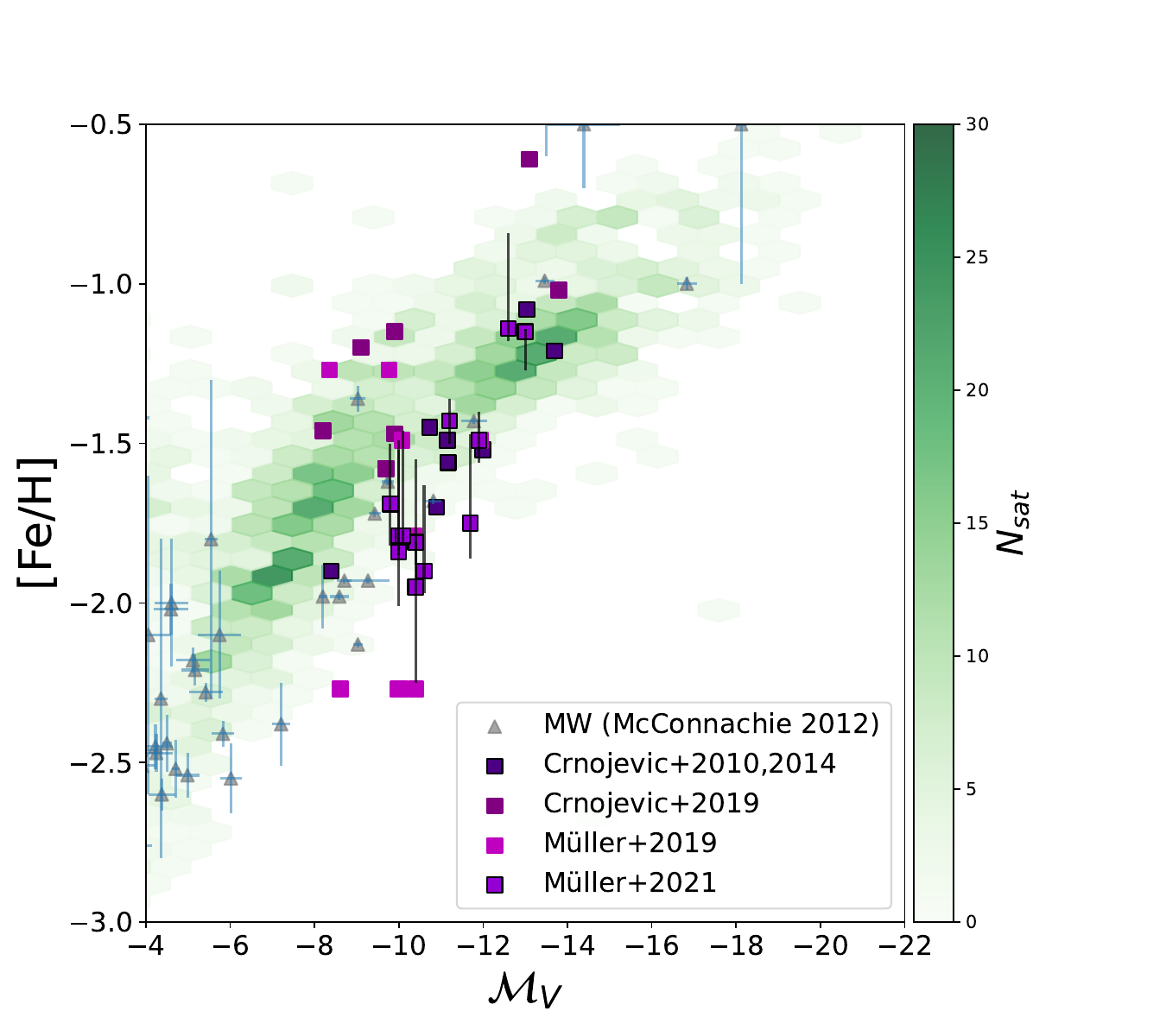}
        \caption{The luminosity--metallicity relation predicted for the Centaurus~A satellites using 30 EPS trees with mass $10^{13} \mathrm{M}_\odot$ (left panel) and N-body trees with satellites within a projected distance of 700~kpc in order to encompass all satellites within the backsplash volume (right panel). The contrast of each hexagon indicates the number of model galaxies falling within that region, as indicated by the color bar to the right of each panel. Observed values are shown by squares \citep{crnojevic2010,crnojevic2019,Muller2019,Muller+2021}. Metallicities of the Milky Way satellites are shown as grey triangles \citep{mcconnachie2012}. The same observed data are reproduced in each panel.}
        \label{fig:feh}
    \end{figure*}

Currently available observed metallicities of Centaurus~A dwarfs are limited in number. In Figure \ref{fig:feh} we show the observations of [Fe/H] for Centaurus~A satellites from \cite{crnojevic2010,crnojevic2014,crnojevic2019} and \cite{Muller2019,Muller+2021}. Notice that \cite{Muller2019} finds a [Fe/H] $\sim -2.25$ metallicity floor between $\mathrm{M}_\mathrm{V}\sim$ $-8$ to $-10$. However, their measurement errors are $\sim 0.5\,\mathrm{dex}$, making the observations consistent with our modeled values. While spectroscopic measurements of metallicities by \cite{Muller+2021} agree well with Milky Way dwarfs, their photometric measurements are reported to have a larger scatter. The authors state that the scatter might be due to the age--metallicity degeneracy and incorrect assumptions of uniformly old ($\sim10$ Gyr) stellar populations. Using the same astrophysical prescriptions and parameters as those that reproduced the luminosity--metallicity relations of the Milky Way dwarfs, we present the modeled metallicities of the Centaurus~A dwarfs in 30 $M_\mathrm{vir}=10^{13}\,\mathrm{M}_{\odot}$ EPS trees and for satellites within 700~kpc for the N-body tree. The metallicities of the modeled dwarfs in this work agree well with currently available observations of Centaurus~A satellites \citep{crnojevic2010,crnojevic2019,Muller2019}. This could potentially mean that the Centaurus~A satellites have a similar enrichment history to that of the Milky Way's satellites and/or that dwarf metallicities are independent of their local environment. In \cite{Weerasooriya+2022}, we show that ram pressure does not significantly affect the luminosity--metallicity relation of the Milky Way satellites. We therefore do not expect a significant change in luminosity--metallicity relation as a function of halo mass. Furthermore, models run with and without ram pressure stripping effects result in very similar luminosity-metallicity relations for the modeled dwarf satellites of Centaurus~A.

\subsection{Velocity Dispersion}

\begin{figure*}[!h]
    \centering
        \includegraphics[clip=true,trim={0cm 0 1cm 2cm},width=.47\textwidth]{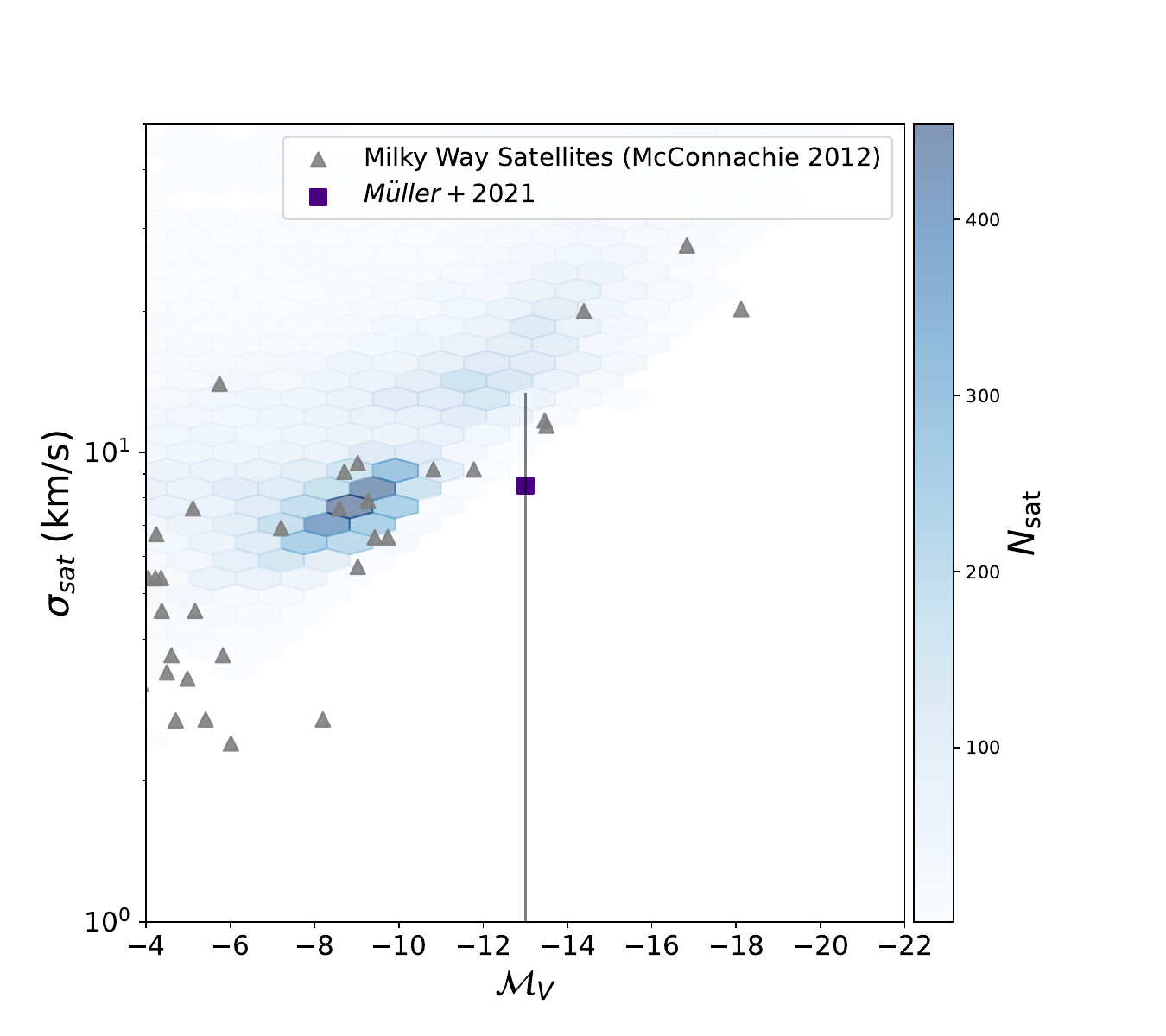}
        \includegraphics[clip=true,trim={0cm 0 1cm 2cm},width=.47\textwidth]{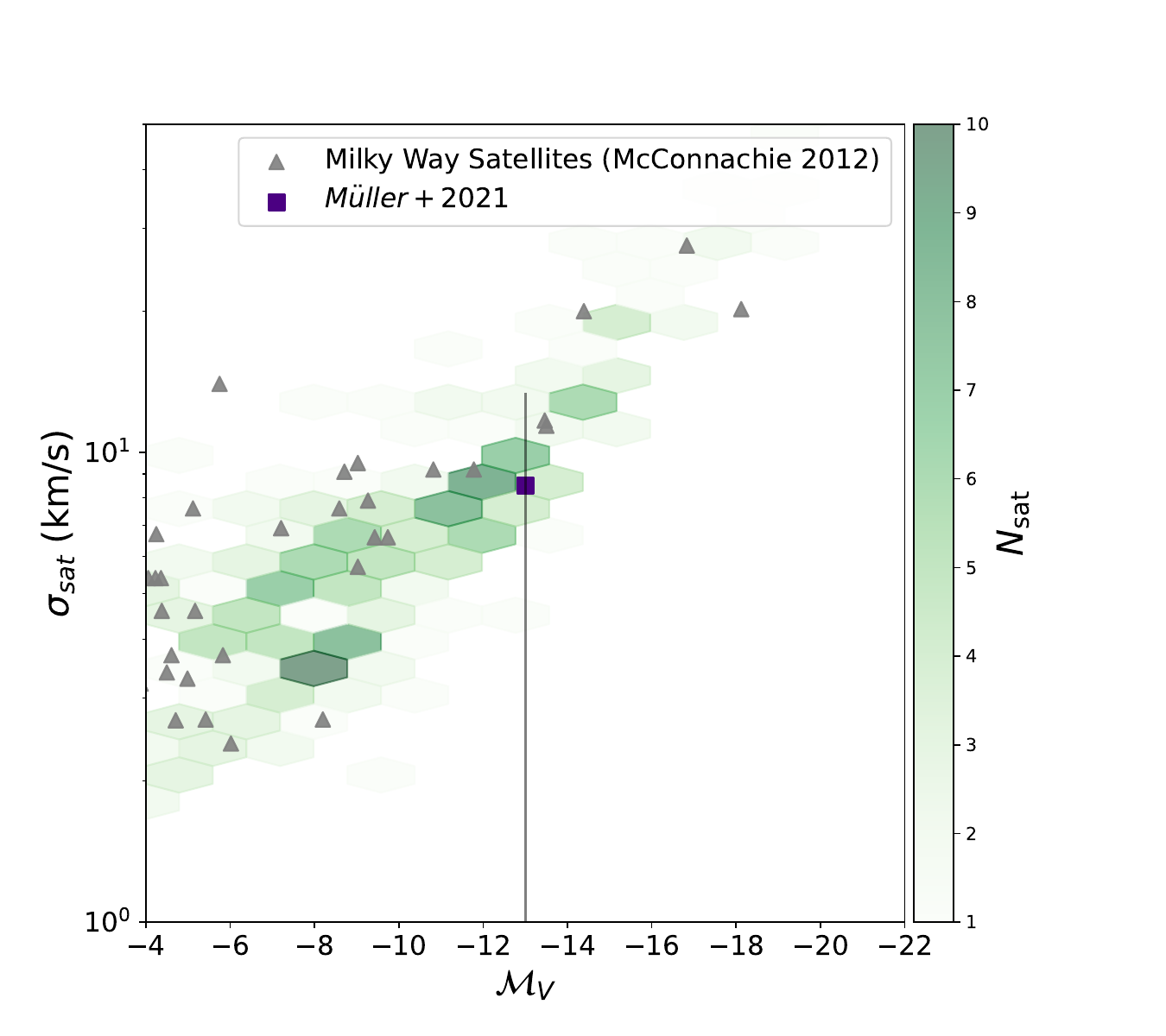}
    \caption{The relation between line-of-sight stellar velocity dispersion (computed at the stellar half-mass radius of each galaxy) and V-band absolute magnitude in Centaurus~A galaxies. We show results for EPS merger trees (left panel) and N-body merger trees (right panel) as blue and green hexagons respectively. For the N-body tree we show all satellites within a projected distance of 700~kpc in order to encompass all satellites within the backsplash volume. The contrast of each hexagon represents the number of model galaxies found within that region, as indicated by the color bar to the right of each panel. Grey triangles show the observed velocity dispersions of the Milky Way satellites \citep{mcconnachie2012}. The purple square indicates the measured velocity dispersion of Centaurus~A satellite galaxy KK197 \protect\citep{Muller+2021}. The same observational data are reproduced in both panels. Note that the predicted values of Centaurus A satellites are consistent with the velocity dispersion of the observed Milky Way satellites.}
    \label{fig:vdis}
\end{figure*}

The velocity dispersion of the stars in a satellite galaxy is largely determined by the gravitational potential of the subhalo in which they live---therefore, velocity dispersions are a sensitive probe of the masses and gravitational potentials of those subhalos \citep{Wake+2012,Bogdan+2015,Schechter2015}. However, velocity dispersions of Centaurus~A satellites are unknown with the exception of KK197. As such, additional data is required before any comparison can be made to the population as a whole. In Figure \ref{fig:vdis}, we show the velocity dispersion of KK197 \citep{Muller+2021}, and those of the Milky Way satellites \citep{mcconnachie2012}. We calculate the velocity dispersion ($\sigma_\mathrm{sat}$) at stellar half-mass radii for the modeled dwarfs. As described in detail in \S\ref{sec:velocityDispersion}, we assume Jeans equilibrium in the total gravitational potential (i.e. self gravity and the gravitational potential of the dark matter halo) when calculating the velocity dispersion of stars as a function of radius $\sigma(r)$. This $\sigma(r)$ is then integrated along the line-of-sight at the half-mass radius, weighted by the luminosity profile of the galaxy to predict the observed velocity dispersion. Most of our modeled sample falls within the observed velocity dispersions of the Milky Way satellites \citep{mcconnachie2012}. Given that velocity dispersion probes dark matter subhalo mass, this suggests that Centaurus~A dwarfs occupy similar halos, at a given luminosity/stellar mass, as their Milky Way counterparts.

\section{Star Formation Histories }

Detailed star formation histories of Centaurus~A dwarfs have not yet been measured observationally. In fact, only a few measurements of the \emph{present day} star formation rates of Centaurus~A dwarfs exist: star formation rates of KK197, ESO-269066, ESO-381018 \citep{Marakova+2007} and five dwarf irregulars KK182 (Cen6), ESO269-58, KK196 (AM1318-444), HIPASS J1348-37, ESO384-16 \citep{Crnojevic+2012} are available in literature. Both KK197 and ESO-269066 are dwarf spheroidals, while ESO-269066 is a dwarf irregular. Dwarf spheroidals typically have old stellar populations whose light is dominated by their red giant branches, while dwarf irregulars are metal-poor and have varying levels of current star formation. \cite{Crnojevic+2012} state that KK197 and ESO-269066 have unusual RGB color scatter, which shows active star formation with high metallicity, while ESO-381018 is a typical dwarf irregular. Two of the dwarf irregulars (KK196 and ESO269-58) studied are within $700$~kpc of Centaurus~A (see Figure 6 of \citealt{Crnojevic+2012}). Positioned in the middle of Centaurus~A's southern radio lobe, KK196 has a star formation rate of $0.0046 \pm 0.0004\,\mathrm{M}_{\odot}\,\mathrm{yr}^{-1}$ and has formed more than $60\%_{-30\%}^{+20\%}$ of its stars more than 5~Gyr ago ($\tau_{60} = 5$~Gyr\footnote{We define $\tau_X$ as the lookback time at which a galaxy had formed $X$\% of its final stellar mass.}; \citealt{Crnojevic+2012}). Meanwhile, ESO269-58, located $300 \pm 50$ kpc from Centaurus~A, has few blue loop, red supergiants and a very broad red giant branch stars, and dense asymptotic giant branch zone. This dwarf has a higher star formation rate compared to KK196 with $0.07 \pm 0.04\,\mathrm{M}_{\odot}\,\mathrm{yr}^{-1}$, and has formed $50\%_{-15\%}^{+15\%}$ of stars more than 5 Gyrs ago ($\mathbf{\tau_{50} = 5}$~Gyr). While its star formation activity has been enhanced between 3-5 Gyrs ago, \cite{Crnojevic+2012} also find that Centaurus~A has lowered its star formation rate in the last 1~Gyr. \\

In this section we provide predictions for these SFHs from our model, along with distributions of quenching times (which we will define below). We remind the reader that the physical prescriptions that determine SFHs (cooling in the circumgalactic medium, star formation rates, feedback models, and ram pressure stripping) are identical to those used by \cite{Weerasooriya+2022}, who demonstrated that these prescriptions could successfully reproduce the SFHs of Milky Way dwarf satellites. We therefore explore the predictions of these prescriptions---which contain the relevant physical scalings to allow them to be extended to the Centaurus~A environment---which may be used to test our model against future observational determinations of SFHs.\\

In addition to full SFHs, we can also infer quenching times---a measure of when the star formation in a galaxy ended. Quenching times have been quantified in a variety of different ways in the literature. Most commonly, the quenching time has been defined as the period for which a satellite remains star forming after infall into its host halo \citep{Rocha+2012,Weisz2015,Foltz+2018}. Other studies define quenching time as the time at which the specific star formation rate falls below $10^{-11}\,\mathrm{yr^{-1}}$ \citep{akins2021} or use both \citep{Baxter+2022}. However, in this work we choose to use the time at which a galaxy had formed 90\% of its present day stellar mass, $\tau_{90}$, as our definition of quenching time. This has the advantage of being observationally-measurable---in \cite{Weerasooriya+2022} we demonstrated that our model produced distributions of $\tau_{90}$ broadly consistent with those measured by \cite{weisz2019} for Milky Way dwarf satellites. 
The distributions of $\tau_{90}$ for Centaurus~A dwarfs will serve as a testable prediction of our model that can be confirmed, or refuted, by future observations.\\

\begin{figure*}[h!]
    \centering
    \includegraphics[width=20cm]{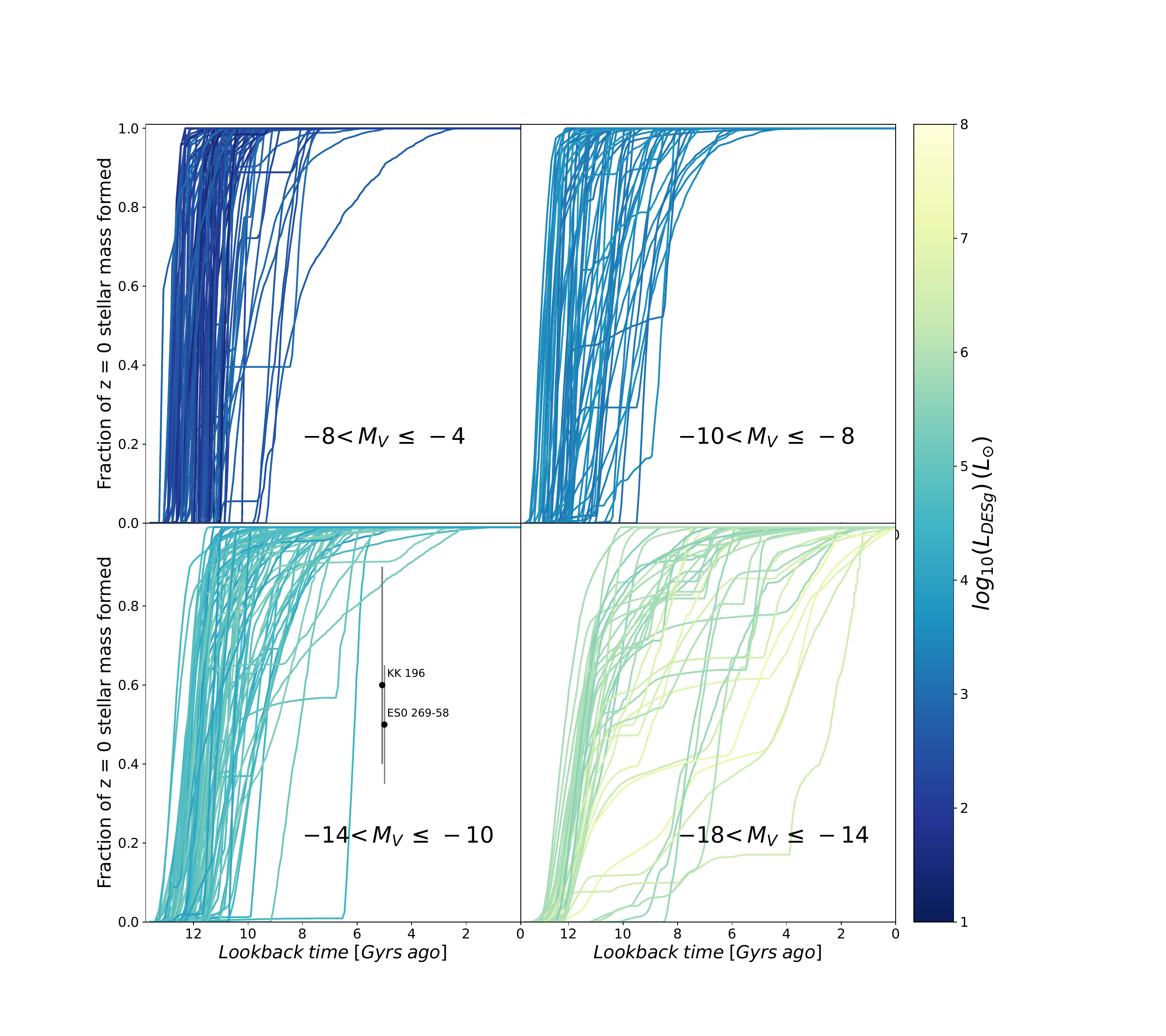}
    \caption{Cumulative star formation histories are shown for Centaurus~A satellites within 700~kpc projected distance from the center of the host halo as predicted by our model using our N-body merger tree. The cumulative fraction of stars formed is shown as a function of look-back time. Individual SFHs are colored by their luminosity in the $\mathbf{g_{DES}}$-band as indicated by the color bar on the right of the figure. Galaxies are divided into four panels based on their absolute magnitudes: $-8 < M_\mathrm{V} \leq -4$, $-10 < M_\mathrm{V} \leq -8$, $-14 <  M_\mathrm{V} \leq -10$, and $-18 < M_\mathrm{V} \leq -14$ as indicated in each panel. In the lower left panel the two available observational measurements are shown as black circles with error bars.}
    \label{fig:SFH}
\end{figure*}

In Figure \ref{fig:SFH}, we show the modeled cumulative star formation histories of the Centaurus~A satellites as a function of look-back time colored by their absolute V-band magnitude at $z\,=\,0$ from our N-body merger tree. As expected, our Centaurus~A modeled SFHs are similar to those of the Milky Way satellites in \cite{Weerasooriya+2022}, indicating no strong influence of host halo environment on the SFHs.\\

\begin{figure*}[h!]
    \centering
    \includegraphics[width=15cm]{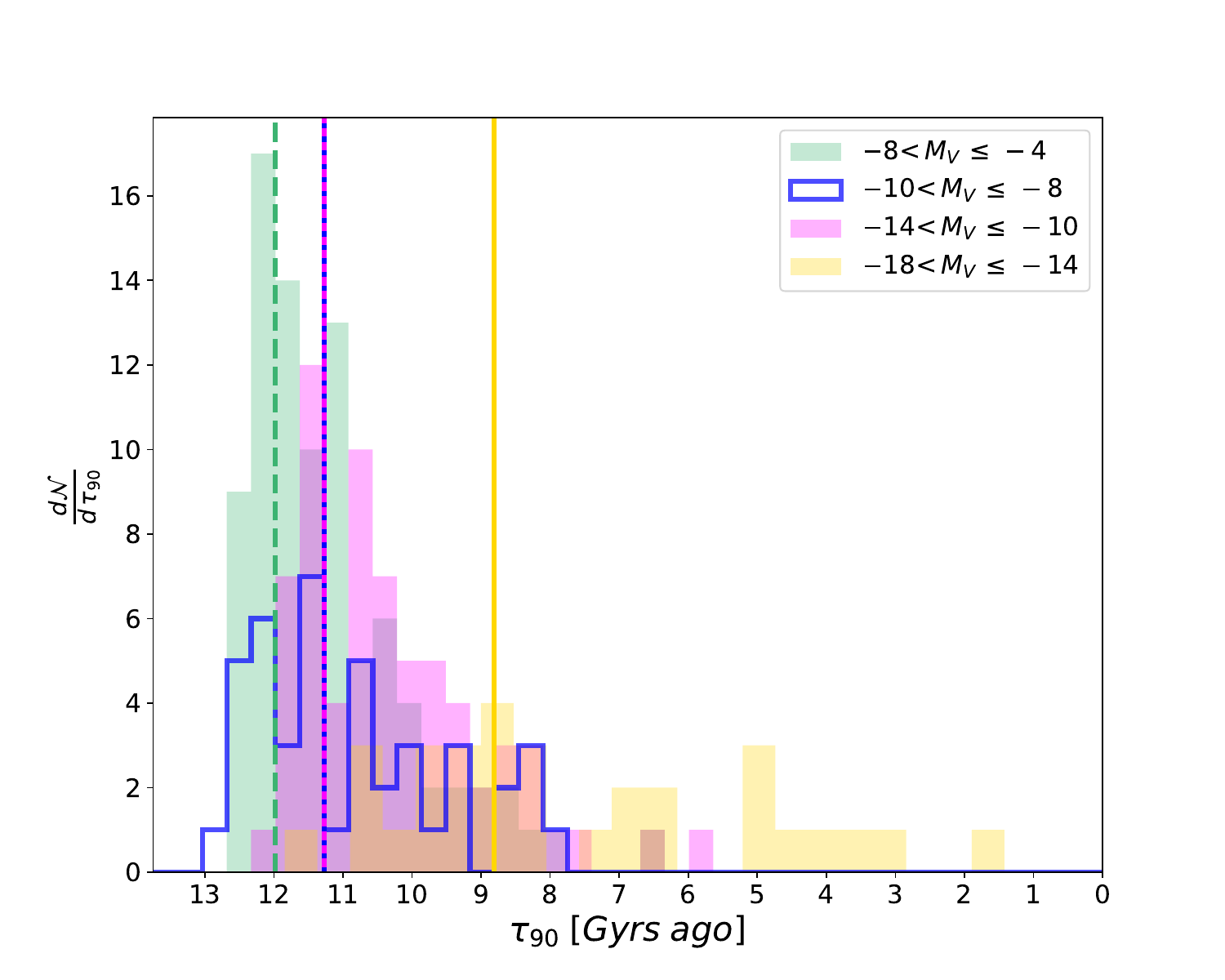}
    \caption{The distributions of quenching times ($\tau_{90}$, the lookback time at which a particular galaxy has formed 90\% of its stellar mass at $z=0$) for the panels shown in Figure \ref{fig:SFH}. The faintest satellites quenched around 12~Gyrs ago, while satellites between $-14\leq M_\mathrm{V}\leq -8$ quenched after 11.2~Gyrs ago. The most luminous galaxies quench much later $\sim 8.8$~Gyrs ago. Median values for each distribution are shown by vertical lines.}
    \label{fig:tau}
\end{figure*}

In the upper left panel, we show the faintest dwarfs with $-8 < M_\mathrm{V} \leq -4$. Most of these ultra-faint dwarfs quenched 8--12 Gyrs ago---this is similar to ultrafaints around the Milky Way which are fossils of the first galaxies \citep{Bovill&ricotti2011,Brown+2012}. While such faint galaxies have not yet been observed in the Centaurus~A system, it is expected that observations will reach these magnitudes in the next generation of surveys. Experiments by \cite{Mutlu-Pakdil+2021} show that LSST could detect Centaurus~A dwarfs down to $M_\mathrm{V}\geq-7$. Furthermore, simulations based on extrapolations of HST observations indicate that the Nancy Grace Roman Space Telescope will be able to resolve stellar populations in M83 which is at a distance of 4.5~Mpc, even further than Centaurus~A \citep{Akeson+2019}.\\

The upper right and lower left panels show satellites in the range $-10 < M_\mathrm{V} \leq -8$ and $-14 < M_\mathrm{V} \leq -10$ respectively. The majority of these satellites quenched 10--12 Gyrs ago, with a small fraction quenching later (6--10 Gyr ago). The two measured $\tau_X$ values for observed Centaurus~A dwarfs discussed earlier fall in the magnitude range of the lower left panel of Figure \ref{fig:SFH}, and are indicated by black circles with error bars. Our model predicts higher stellar mass fractions formed by these times than the observations suggest, although the uncertainties in the measurements remain large. The brightest galaxies  ($-18 < M_\mathrm{V} \leq -14$), shown in the lower right panel, show noticeably more extended SFHs. This is clearly seen in Figure \ref{fig:tau} which shows the corresponding distributions of $\tau_{90}$ for the same four intervals of absolute magnitude.

\section{Summary \& Conclusions}\label{sec:conc}

We have presented predictions for the properties of the dwarf galaxy population orbiting the Centaurus~A galaxy, using our semi-analytic model \textsc{Galacticus} applied to merger trees produced from both an N-body simulation and from application of extended Press-Schechter theory. This model was previously calibrated to match the observed properties of Milky Way dwarf galaxies \citep{Weerasooriya+2022}, and we retain the same calibration here, allowing us to make testable predictions for the Centaurus~A system. Importantly, the astrophysics models controlling key baryonic processes (e.g. star formation, feedback, and ram pressure stripping) incorporate the expected scalings of these processes with host halo mass and environment. As such, the predictions of our model for the dwarf population of the Centaurus~A system can be seen as a potential test of the underlying physics. We find that our model is able to reproduce the overall properties of the Centaurus~A dwarf population quite well, given the sparse nature of the current observations. \\

In particular, we find that our model is able to correctly reproduce the number of dwarf galaxies found within a projected radius of 150~kpc of Centaurus~A (the region for which observational surveys are reasonably complete), although it predicts somewhat more of the brightest satellite galaxies than are observed in this inner region. Expanding to the full extent of the Centaurus~A system (i.e. out to the expected splashback radius of the system at around 700~kpc) we find that our model is in good agreement with the number of known highly luminous satellites. This suggests that the excess of bright satellite galaxies in the inner 150~kpc region predicted by our model may simply be a result of random fluctuations in galaxy positions, as we have only a single N-body realization of this system. Further testing of this hypothesis is planned using additional N-body simulations, and the more detailed orbital model of satellite galaxies employed by \cite{2023arXiv230813599A}. At fainter luminosities, our model predicts between 200 and 500 satellites brighter than $M_\mathrm{M} = -4$ within the backsplash radius. Compared to the approximately 120 known dwarfs (down to the current detection threshold of  $M_\mathrm{M} = -8$) this suggests that many hundreds of dwarf satellites remain to be discovered around Centaurus~A.\\

In addition to these global properties, we have also explored predictions for internal properties of Centaurus~A dwarfs. We find that our model predicts half-light radii of these dwarfs which are in good agreement with observations for brighter ($M_\mathrm{M} < -10$) dwarfs, but overpredicts sizes of fainter systems. A similar conclusion was reached for Milky Way dwarfs by \cite{Weerasooriya+2022}, who suggest that this may be due to the limited resolution of the associated N-body halos, an inaccuracies in the extended Press-Schechter approach for lower halo masses.\\

We find that our model predicts metallicities as a function of luminosity that are consistent with observations in the Centaurus~A system. Given that, in our model, the metallicity attained by a galaxy is largely driven by the process of stellar feedback (which ejects gas and metals from galaxies), and halo assembly history, this suggests that our models of these processes are consistent with observations. Furthermore, we find that the metallicity--luminosity ratio of model dwarfs in the Centaurus~A system is consistent with that predicted for the Milky Way dwarfs by \cite{Weerasooriya+2022}. This implies that environmental processes (such as ram pressure stripping), which may affect metal enrichment in our model, are sub-dominant to internal processes (star formation and feedback). To further test this hypothesis we have repeated our model calculations without the effects of ram pressure stripping, finding almost no change in the predicted metallicity--luminosity relation, consistent with the conclusions of the observational analysis of \cite{Taibi+2022}.\\

Predicted line-of-sight velocity dispersions of model galaxies as a function of luminosity are also found to be consistent with those found by \cite{Weerasooriya+2022} for Milky Way dwarfs. As velocity dispersion in dwarfs is largely determined by the depth of the gravitational potential well of the dwarf's subhalo, this implies that dwarfs of a given luminosity form in subhalos of approximately the same mass independent of environment. The single measured velocity dispersion in the Centaurus~A system (for dwarf galaxy KK197) is consistent with our model predictions (although the measurement is highly uncertain).\\

Finally, we have also explored predictions for the star formation histories (SFHs) of Centaurus~A dwarfs, characterizing these by the cumulative mass fraction of stars formed as a function of lookback time, and by the summary statistic $\tau_{90}$ (the lookback time at which 90\% of the final stellar mass had formed). Once again, our results closely resemble those predicted for Milky Way dwarfs by \citeauthor{Weerasooriya+2022}~(\citeyear{Weerasooriya+2022}; which were themselves consistent with observational estimates of SFHs in those systems), also indicating that internal processes are dominant in guiding the formation of these dwarf galaxies. Observational measures of star formation histories exist for only two Centaurus~A dwarfs presently. While these measurements are highly uncertain, they suggest star formation delayed relative to the predictions from our model. Future measurements may therefore be able to strongly test and rule out the specific star formation model adopted in this work.\\

Dwarf galaxies, with their shallow potential wells and large numbers, are uniquely well-suited to testing models of feedback and environmental effects in galaxy formation. Given the large theoretical uncertainties in these processes, comparing models to observations across a range of environments can provide a strong and useful test of whether internal or external processes are responsible for shaping the evolution of these galaxies. Using our current model, presented in \cite{Weerasooriya+2022}, this work shows that internal processes (star formation and feedback) seem to be the dominant drivers of dwarf galaxy evolution. While the current observations in the Centaurus~A system are largely consistent with our predictions, the limited number and precision of these observations do not currently provide strong tests of this model. However, there are hints of tensions between theory and observations (for example, in the histories of star formation in more luminous dwarfs). The next generation of galaxy surveys (to be carried out by facilities such as the Rubin Observatory and Nancy Grace Roman Space Telescope) have the potential to dramatically increase both the size and precision of observational samples. The predictions presented in this work are likely to be strongly testable in the near future.\\

\noindent We thank the anonymous referee for their valuable feedback and insightful suggestions, which have significantly enhanced the quality of this manuscript. The authors acknowledge the University of Maryland supercomputing resources ({\tt http://hpcc.umd.edu}) made available for conducting the research reported in this paper. \\
% \newpage
% \textbf{NOTES TO THE REFEREE}

% \textbf{Please note that we include here two figures as a response to the referee report for referee's comment number 24.}

% \begin{figure}[h!]
%     \centering
% \includegraphics[width=.45\textwidth]{FeH_Nbody.pdf}
% \includegraphics[width=.45\textwidth]{FeH_Nbody_ram0.jpg}
%     \caption{Here we show two models for N-body trees with 100\% ram pressure stripping efficiency and 0\% efficiency. While the number of galaxies in each bin does change, the overall trend in luminosity-metallicity is not affected by the change in strength of ram pressure stripping.
% }
%     \label{fig:response}
% \end{figure}
\noindent \textit{Software:} \textsc{Galacticus} \citep{galac}, \texttt{ROCKSTAR} \citep{rockstar}, \texttt{AMIGA} \citep{AHF}, \texttt{CONSISTENT\_TREES} \citep{consistent_trees}, \texttt{JUPYTER} \citep{Jupyter}, \texttt{NUMPY} \citep{numpy}, \texttt{SCIPY} \citep{scipy}, and MATPLOTLIB \citep{matplotlib}.

\bibliography{sample631,ref2.bib,ref}{}
\bibliographystyle{aasjournal}



\end{document}